\documentclass[aps,nofootinbib,showkeys,noshowpacs,preprintnumbers,amsmath,amssymb]{revtex4}
\pdfoutput=1

\def\be{\begin{equation}}
\def\ee{\end{equation}}
\def\ba{\begin{eqnarray}}
\def\ea{\end{eqnarray}}
\def\bea{\begin{eqnarray}}
\def\eea{\end{eqnarray}}
\def\bes{\begin{subequations}}
\def\ees{\end{subequations}}

\newcommand{\sm}{{\sigma_{\rm m}}}

\newcommand{\tk}{{\widetilde k}}
\newcommand{\tK}{{\widetilde K}}

\newcommand{\A}{{\mathcal{A}}}

\newcommand{\ta}{{\widetilde a}}
\newcommand{\tal}{{\widetilde \alpha}}

\newcommand{\td}{{\widetilde d}}

\newcommand{\MSbar}{\overline{\rm MS}}

\usepackage{graphics}
\usepackage{graphicx}
\usepackage{dcolumn}
\usepackage{bm}
\usepackage{epsfig}
\usepackage{graphicx,color}
\usepackage{multirow}

\begin{document}


\title{Borel-Laplace Sum Rules with $\tau$ decay data, using OPE with improved anomalous dimensions}

\author{C\'esar Ayala$^a$}
\email{c.ayala86@gmail.com}
\author{Gorazd Cveti\v{c}$^b$}
\email{gorazd.cvetic@gmail.com}
\author{Diego Teca$^b$}
\email{diegotecawellmann@gmail.com}
\affiliation{$^a$Instituto de Alta Investigaci\'on, Sede Esmeralda, Universidad de Tarapac\'a, Av.~Luis Emilio Recabarren 2477, Iquique, Chile}
\affiliation{$^b$Department of Physics, Universidad T{\'e}cnica Federico Santa Mar{\'\i}a, Avenida España 1680, Valpara{\'\i}so, Chile}

\date{\today}

\begin{abstract}
  We perform numerical analysis of double-pinched Borel-Laplace QCD sum rules for the strangeless semihadronic $\tau$-decay data. The $D=0$ contribution to the theoretical contour integral in the sum rules is evaluated by the (truncated) Fixed Order perturbation theory method (FO) and by the Principal Value (PV) of the Borel integration. We use for the full Adler function the Operator Product Expansion (OPE) with the terms $\sim \langle O_D \rangle$ of dimension $D=2 n$ where $2 \leq n \leq 5$ for the (V+A)-channel, and $2 \leq n \leq 7$ for the V-channel data. In our previous works \cite{EPJ21,EPJ22}, only the (V+A)-channel data was analysed. In this work, the analysis of a new set of V-channel data is performed as well. Further, a renormalon-motivated construction of the $D=0$ part of the Adler function is improved in the $u=3$ infrared renormalon sector, by involving the recently known information on the two principal noninteger values $k^{(j)}=\gamma^{(1)}(O_6^{(j)})/\beta_0$ of the effective leading-order anomalous dimensions. Additionally, the OPE of the Adler function has now the D=6 contribution with the principal anomalous dimension ($\sim \alpha_s^{k^{(1)}}$), and terms of higher dimension (with zero anomalous dimension). Cross-checks of the obtained extracted values of $\alpha_s$ and of the condensates were performed by reproduction of the (central) experimental values of several double-pinched momenta $a^{(2,n)}$. The averaged final extracted values of the ($\MSbar$) coupling are: $\alpha_s(m_{\tau}^2) = 0.3169^{+0.0070}_{-0.0096}$, corresponding to $\alpha_s(M_Z^2)=0.1183^{+0.0009}_{-0.0012}$.
  \end{abstract}
\keywords{perturbative QCD; QCD phenomenology; semihadronic $\tau$ decays; renormalons}

\maketitle

\section{Introduction}
\label{sec:Int}

One of the central problems of QCD is to determine the value of the QCD running coupling $a(Q^2) \equiv \alpha_s(Q^2)/\pi$, usually defined in the $\MSbar$ scheme \cite{MSbar}. The coupling $a(Q^2)$ depends on the squared momentum $Q^2 \equiv - q^2$ [$= - (q^0)^2 + {\vec q}^2$] for real spacelike momenta ($Q^2>0$). This coupling can be considered in the generalised spacelike (Euclidean) regime, i.e., for complex $Q^2$ with the exception of the negative (timelike) values $Q^2 = -s < 0$. In perturbative QCD (pQCD), the $\MSbar$ coupling has $Q^2$-dependence as governed by the known (five-loop) $\MSbar$ $\beta$-function $\beta(a)$; this specific polynomial (i.e., perturbative) form of $\beta(a)$ then in turn results in a running coupling $a(Q^2)$ which has singularities not just for timelike $Q^2 < 0$, but also in a Euclidean regime $0 \leq Q^2 \leq \Lambda^2_{\rm Lan}$, where $\Lambda^2_{\rm Lan} \sim 0.1 \ {\rm GeV}^2$, and $[0,\Lambda^2_{\rm Lan}]$ represents the region of the Landau (cut) singularities.\footnote{These singularities in $a(Q^2)$ in pQCD do not reflect the behaviour of spacelike observables ${\cal D}(Q^2)$ such as DIS structure functions, or quark current correlators (and the related Adler function). Namely, spacelike QCD observables ${\cal D}(Q^2)$ are holomorphic functions of $Q^2$ in the entire generalised spacelike regime $Q^2   \in \mathbb{C} \backslash (-\infty, 0)$, which is a consequence of the general principles of Quantum Field Theory \cite{Bogolyubov:1959bfo,Oehme:1994pv}. A large class of pQCD schemes where Landau singularities are not on the real axis in the $Q^2$-plane was investigated in Ref.~\cite{LandCCO}.} These singularities indicate that pQCD has problems in describing the physics at low momenta ($|Q^2| \lesssim 1 \ {\rm GeV}^2$).

The semihadronic decays of the $\tau$-leptons have been well measured by OPAL \cite{OPAL,PerisPC1} and ALEPH \cite{ALEPH2,DDHMZ,ALEPHfin,ALEPHwww} Collaborations, and the precision of the results of the ALEPH Collaboration is high. The semihadronic $\tau$-lepton decays probe QCD effects at relatively low momenta, $|Q^2| \lesssim m^2_{\tau} \approx 3 \ {\rm GeV}^2$, where, as mentioned above, pQCD approaches start having problems due to the growth of the coupling $a(Q^2)$ and the vicinity of the Landau singularities. This theoretical problem, and the mentioned high precision of the experimental ALEPH data, imply that the theoretical analysis of the semihadronic $\tau$-decays is an important, and challenging, task at the frontier of validity of pQCD. The central theoretically relevant spacelike quantity in the physics of strangeless semihadronic $\tau$-decays is the $u$-$d$ quark current correlator (polarisation) function $\Pi(Q^2)$ and the related Adler function ${\cal D}(Q^2) \propto d \Pi(Q^2)/d \ln Q^2$. The OPE is believed to describe reasonably well the behaviour of the Adler function ${\cal D}(Q^2)$ in this frontier regime between the intermediate and low momenta, $|Q^2| \approx 3 \ {\rm GeV}^2$, and it contains relevant contributions of dimension $D=0, 4, 6$ and possibly higher ($D=2$ contributions are chiral and are negligible due to the smallness of the quark masses $m_u$ and $m_d$). On the other hand, in the $D=0$ contribution $d(Q^2)_{D=0}$ of the Adler function, the terms $D=4,6,\ldots$ are reflected in the IR renormalon singularities at $u=2,3,\ldots$ \cite{ren}. The strength of these singularities is related to the $Q^2$-dependence of the corresponding OPE terms $D=4,6,\ldots$, such that the IR renormalon ambiguities of the $D=0$ contribution can be cancelled by the corresponding $D > 0$ OPE terms. 

The bases of the theoretical description of the semihadronic $\tau$-decays were developed and applied in the works \cite{NP88,B88B89,BNP92,DP92,SEW,dpEW}. The perturbation expansion of the $D=0$ contribution $d(Q^2)_{D=0}$ to the Adler function has been calculated up to ${\cal O}(a^4)$ \cite{d1,d2,BCK}. Furthermore, presumably good estimates \cite{KatStar,BCK,Boitoetal} are known also for the coefficient at $a^5$ in the expansion of $d(Q^2)_{D=0}$. 

In our previous works \cite{EPJ21,EPJ22} we applied double-pinched Borel-Laplace sum rules to the ALEPH data for the strangeless (V+A)-channel semihadronic $\tau$-decays, where V and A are short notations for vector and axial vector, respectively. We used for $D=0$ Adler function contribution $d(Q^2)_{D=0}$ as the basis the Borel transform ${\cal B}[\td](u)$ of a related auxiliary function $\td(Q^2)$ (cf.~Ref.~\cite{renmod}) which agrees with $d(Q^2)_{D=0}$ only at the one-loop level. In our previous works \cite{EPJ21,EPJ22,Trento} we used for ${\cal B}[\td](u)$ an expression which contains the correct structure of the $u=2$ infrared (IR) renormalon, and the large-$\beta_0$ structure of the $u=3$ IR renormalon, namely the terms $\sim 1/(3-u)^2, 1/(3-u)$. The transform ${\cal B}[\td](u)$ also contains the leading $u=-1$ ultraviolet renormalon term $\sim 1/(1+u)^2$. In the work \cite{EPJ21} the OPE expansion was taken with $D \geq 4$ terms all with zero anomalous dimensions and was truncated either at $D=8$ or $D=10$ term. In \cite{EPJ22,Trento}, the OPE was taken with two $D=6$ terms which reflected the mentioned $u=2$ large-$\beta_0$ IR renormalon structure, and truncated there.

The present work represents an improvement on several aspects of our mentioned previous works \cite{EPJ21,EPJ22,Trento}. In the mentioned Borel transform ${\cal B}[\td](u)$ we refine the $u=3$ IR renormalon terms, by incorporating the approximate noninteger (i.e., beyond large-$\beta_0$) powers of the $u=3$ IR renormalon contributions, ${\cal B}[\td](u) \sim 1/(3 -u)^{\kappa^{(1)}}, 1/(3 -u)^{\kappa^{(2)}}$, where $\kappa^{(1)}=0.778$ and $\kappa^{(2)}=0.375$, as implied by the work \cite{BHJ} (cf.~also \cite{JK,Lanin:1986zs,ACh}) on the anomalous dimensions of $D=6$ operators in the (V+A)-channel. We note that the values of these power indices differ significantly from the corresponding mentioned large-$\beta_0$ (LB) values $\kappa^{(1)}_{\rm LB}=2$ and $\kappa^{(2)}_{\rm LB}=1$. We have the theoretical requirement that the $Q^2$-dependence in $d(Q^2)_{D=0}$ generated by these terms is the same as the $Q^2$-dependence in the $D=6$ OPE terms of the full Adler function $d(Q^2)$, which enables the so called renormalon cancellation in the OPE. In principle, the corresponding $D=6$ OPE contribution to the Adler function in the (V+A)-channel is
\be
d(Q^2)_{D=6} = \frac{6 \pi^2}{(Q^2)^3}  \left[ \langle O^{(1)}_{6} \rangle_{\rm V+A} \;  a(Q^2)^{k^{(1)}} + \langle O^{(2)}_{6} \rangle_{\rm V+A} \;  a(Q^2)^{k^{(2)}} \right],
\label{dD6} \ee
where the anomalous dimensions $k^{(j)}=\gamma^{(1)}(O_6^{(j)})/\beta_0$ ($j=1,2$) are related to the aforementioned power indices $\kappa^{(j)}$ by the simple relation $\kappa^{(j)}=1-k^{(j)}$ (i.e., we have $k^{(1)} = 0.222$ and $k^{(2)}=0.625$). This relation is a consequence of the mentioned requirement of the same $Q^2$-dependence of the $D=6$ contributions and of the IR renormalon $u=3$ contribution in the $D=0$ part. We note that the value $k^{(1)} = 0.222$ was obtained by taking the arithmetic average of the first two smallest anomalous dimensions obtained in \cite{BHJ} (the two values are close to each other, $\gamma^{(1)}(O_6^{(1)}) \approx 0.5$), and the value $k^{(2)}=0.625$ from the next two smallest anomalous dimensions (also close to each other, $\gamma^{(1)}(O_6^{(2)}) \approx 1.4$).\footnote{The leading $\beta$-function coefficient is $\beta_0=9/4$ for the considered case $n_f=3$.} Altogether, there are nine $D=6$ nonchiral operators for the (V+A)-channel Adler function, but the existence of the terms with the higher anomalous dimensions  $\gamma^{(1)}(O_6^{(j)})$ (and thus lower power indices $\kappa^{(j)}$) are neglected in ${\cal B}[\td](u)$. We also note that the $D=4$ operator (gluon condensate) has $\gamma^{(1)}(O_4)=0$.

In practice, when fitting the OPE of Borel-Laplace sum rules, it will turn out that the inclusion of the two terms Eq.~(\ref{dD6}) in the OPE will lead to strong cancellation between these two $D=6$ contributions and, consequently, to numerical instabilities when we include higher dimensional terms $D >6$ in the OPE because in such a case the mentioned cancellation becomes even stronger. Therefore, we will take for $D=6$ contribution to the OPE only the first term in Eq.~(\ref{dD6}). Further, the contributions with other dimensions will all be taken with zero anomalous dimension (for the terms $D > 6$ we do not know their theoretical values).

Another extension, in comparison to our previous works \cite{EPJ21,EPJ22,Trento}, will be the inclusion of the analysis of the V-channel. For that channel, we will use the data of Refs.~\cite{Boito:2020xli,Perisetal}, which combine the V-channel data of ALEPH and OPAL and is supplemented by the $e^+ e^- \to {\rm hadrons}$ cross-section data for some specific exclusive modes, and recent BABAR results for the $\tau \to K^- K^0 \nu_{\tau}$ decay \cite{Boito:2020xli,Perisetal}. It has been known \cite{BNP92} that the number of OPE terms needed for reasonable extracted results is higher in the V-channel than in the (V+A)-channel. In our approach with Borel-Laplace tranforms, it turns out that we will need to include OPE terms up to $D=10$ for the (V+A)-channel, and up to $D=14$ for the V-channel.

Yet another improved aspect, in comparison to the analyses in Refs.~\cite{EPJ21,EPJ22,Trento}, is a better evaluation of the experimental uncertainties of the extracted values of parameters, i.e., of $\alpha_s$ and of the condensates. In our previous works we roughly estimated these uncertainties by applying the approximation that the Borel-Laplace sum rules at different scales are not correlated. It turns out that the correlations are strong, and the application of the systematic evaluation of the correlated experimental uncertainties, e.g. by the method described in the Appendix of Ref.~\cite{Bo2011}, gives us a significantly larger estimate of the experimental uncertainties; in the (V+A)-channel, though, they still remain significantly smaller than the theoretical uncertainties.

The duality violation (DV) effects are strongly suppressed in our sum rules, as will be argued later, by the weight functions that are double-pinched in the Minkowskian regime and have an additional exponential suppression in that regime.

The theoretical part of the sum rules is represented by a contour integral in the complex $Q^2$-plane, with radius $|Q^2| = {\sigma}_{\rm max}$ which is the upper limit of the highest well-measured bin where $\sigma$ denotes the square of the invariant mass of the (strangeless) hadronic product in the measured $\tau$-decays. In the (V+A)-channel ALEPH data, ${\sigma}_{\rm max} \equiv \sm = 2.8 \ {\rm GeV}^2$; in the V-channel data $\sm=3.057 \ {\rm GeV}^2$. The integrand in the sum rule contour integral includes as a factor the full Adler function on the contour, ${\cal D}(\sm e^{i \phi})$, and thus also the $D=0$ contribution $d(\sm e^{i \phi})_{D=0}$ of the Adler function. The corresponding $D=0$ contribution of the contour integral can be evaluated by various methods; we employ the (truncated) Fixed Order perturbation theory (FO), and the Borel integration with the Principal Value renormalon-ambiguity fixing (PV). Both methods involve truncation with a truncation index $N_t$ in the $D=0$ sector: (I) The FO method evaluates the $D=0$ contribution to the sum rule integral as a sum of powers of $a(\sm)$ truncated at $a(\sm)^{N_t}$. (II) The PV method evaluates the $D=0$ part $d(\sm e^{i \phi})_{D=0}$ of the Adler function as the Principal Value of the Borel integration of the most singular part of the Borel transform, ${\cal B}[d](u)_{\rm sing}$, and adds a ``correction'' polynomial $\delta d(\sm e^{i \phi})^{[N_t]}_{D=0}$ truncated at $a(\sm)^{N_t}$, where the latter polynomial is largely free of the renormalon effects; subsequently, the sum of these two contributions is then used in the ($D=0$ part of the) sum rule contour integral.

In principle, we also have the option to apply the Contour Improved perturbation theory (CI) method \cite{CI1,CI2,CIAPT} to the sum rules, i.e., to integrate each power $a(\sm e^{i \phi})^n$ in the integrand (in the $D=0$ sum rule contour integral) numerically, with $a(\sm e^{i \phi})$ running along the contour according to the (five-loop $\MSbar$) Renormalisation Group Equation (RGE). However, as argued in \cite{Hoang1} (cf.~also \cite{Hoang2,Hoang3}), the (truncated) CI approach is inconsistent with the standard treatment of the OPE. Furthermore, the CI approach does not respect the renormalon structure of the considered sum rules when truncated \cite{Hoang1,BJ,BJ2,EPJ21}. It is expected to give numerically extracted results significantly different from those of the methods FO and PV. We will not apply the CI approach in our work.

In our approach with the (double-pinched) Borel-Laplace sum rules, we extract at each truncation index $N_t$ the corresponding values of $\alpha_s$ and of the condensates in the two methods (FO and PV). The optimal value of $N_t$, in each method, is then fixed by requiring that the (double-pinched) momenta sum rules $a^{(2,n)}(\sm)$ [$0 \leq n \leq 2$ for (V+A)-channel; $0 \leq n \leq 4$ for V-channel] are locally insensitive to the variation of $N_t$ around such value of $N_t$.

In the paper, in Sec.~\ref{sec:SRgen} we discuss the (truncated) OPE for the Adler function that will be used in the theoretical part of the sum rules, summarise the sum rules approach for the semihadronic $\tau$-decays, and specify the weight functions of the specific considered sum rules. In Sec.~\ref{sec:renmod} we present the renormalon-motivated model for the $D=0$ part of the Adler function, $d(Q^2)_{D=0}$, introduce the corresponding auxiliary ($D=0$) function $\td(Q^2)$ and its Borel transform ${\cal B}[\td](u)$, and we derive from there the expansion of the Borel transform ${\cal B}[d](u)$ of $d(Q^2)_{D=0}$ around its singularities. In that Section we also present the resulting characteristic function $G_d(t)$ of $d(Q^2)_{D=0}$. In Sec.~\ref{sec:fitBL} we present the results of the numerical analysis, which includes the fits with the Laplace-Borel sum rules, of the (V+A)-channel ALEPH data in Sec.~\ref{subs:VA}, and of the V-channel combined data in Sec.~\ref{subs:V}. In addition, in Sec.~\ref{subs:Nt} we present the $N_t$-dependence of the sum rule momenta $a^{(2,n)}(\sm)$. In Sec.~\ref{sec:concl} we present our final results for the value of the coupling $\alpha_s(m^2_{\tau})$ [and $\alpha_s(m^2_Z)$], i.e., we average the results for $\alpha_s$ of the previous Section over the two considered evaluation methods (FO and PV), and subsequently average these results over the (V+A)-channel and V-channel case. In addition, in Sec.~\ref{sec:concl} we compare the obtained results with our previous results \cite{EPJ21,EPJ22} and the results obtained by other groups from the semihadronic $\tau$-decays. In Appendices \ref{appUV1}-\ref{appGd} we present some additional technical details related with Sec.~\ref{sec:renmod}. Appendix \ref{apperr} is related with Sec.~\ref{sec:fitBL} and contains some technical details about the experimental uncertainties of the extracted parameter values.

\section{Sum rules and Adler function}
\label{sec:SRgen}

In our analysis, the Adler function will play the central role. Its origins are in the quark current correlator
\be
\Pi_{\rm{J}, \mu\nu}(q) =  i \int  d^4 x \; e^{i q \cdot x} 
\langle T J_{\mu}(x) J_{\nu}(0)^{\dagger} \rangle
=  (q_{\mu} q_{\nu} - g_{\mu \nu} q^2) \Pi_{\rm J}^{(1)}(Q^2)
+ q_{\mu} q_{\nu} \Pi_{\rm J}^{(0)}(Q^2).
\label{PiJ}
\ee
Here, $J_{\mu}$ are the up-down quark vector and axial vector charged currents, $J_{\mu} = {\bar u} \gamma_{\mu} d$ and ${\bar u} \gamma_{\mu} \gamma_5 d$ for $J=V, A$, respectively. As mentioned earlier, $q^2 \equiv -Q^2$ is the square of the momentum transfer, $q^2=(q^0)^2 - {\vec q}^2$. We consider first the (V+A)-channel, and thus the corresponding polarisation function $\Pi(Q^2)$ of the quark current correlator is
\be
\Pi(Q^2)_{\rm V+A} = \Pi_{\rm V}^{(1)}(Q^2) + \Pi_{\rm A}^{(1)}(Q^2) + \Pi_{\rm A}^{(0)}(Q^2).
\label{Pidef}
\ee
The term $\Pi_{\rm V}^{(0)}(Q^2)$ is not included because it gives negligible contribution to the sum rules since ${\rm Im} \Pi_{\rm V}^{(0)}(-\sigma + i \epsilon) \propto (m_d-m_u)^2$. Furthermore, the quark mass corrections ${\cal O}(m^2_{u,d})$ and  ${\cal O}(m^4_{u,d})$ are also numerically negligible and will not be included.

The Adler function ${\cal D}(Q^2)$ is the logarithmic derivative of the quark current polarisation function $\Pi(Q^2)$
\be
{\cal D}(Q^2)_{\rm V+A} \equiv - 2 \pi^2 \frac{d \Pi(Q^2)_{\rm V+A}}{d \ln Q^2}.
\label{Ddef}
\ee
In contrast to the polarisation function $\Pi(Q^2)$, the Adler function ${\cal D}(Q^2)$ has no dependence on the renormalisation scale $\mu^2$; it is also renormalisation scheme independent, i.e., it is a (spacelike) observable.

The theoretical expression for the polarisation function $\Pi(Q^2)$ has an OPE form \cite{SVZ}. The OPE form for $\Pi(Q^2)$ usually used in the literature in the numerical analyses (cf.~\cite{Pich,Pich3,RodS,Bo2011,Bo2012,Bo2015,Bo2017,EPJ21}) is
\be
\Pi_{\rm th}(Q^2; \mu^2)_{\rm V+A}  = - \frac{1}{2 \pi^2} \ln \left( \frac{Q^2}{\mu^2} \right) + \Pi(Q^2)_{D=0} + \sum_{p \geq 2} \frac{ \langle O_{2 p} \rangle_{\rm V+A} }{(Q^2)^p} \left( 1 + {\cal O}(a) \right).
\label{PiOPE1}
\ee
Here, $\langle O_{2 p} \rangle_{\rm V+A}$ are effective condensate (vacuum expectation) values of operators of dimension $D=2 p$ ($\geq 4$), for the full channel V+A. The corresponding OPE of the Adler function (\ref{Ddef}) is
\be
{\cal D}_{\rm th}(Q^2)_{\rm V+A}  \equiv  - 2 \pi^2 \frac{d \Pi_{\rm th}(Q^2)}{d \ln Q^2} = 1 + d(Q^2)_{D=0} + 2 \pi^2 \sum_{p \geq 2} \frac{p \langle O_{2 p} \rangle_{\rm V+A} }{ (Q^2)^p}.
\label{DOPE1}
\ee
However, for $D (\equiv 2 p) \geq 6$ terms, the OPE expansion (\ref{PiOPE1}) [$\leftrightarrow$ (\ref{DOPE1})] is not exactly correct. Namely, for each dimension $D = 2 p$ there are in general various $D$-dimensional operators $O_D^{(j)}$ which are accompanied in the OPE by powers $a^{k_D^{(j)}}$ where $k_D^{(j)} = \gamma^{(1)}(O_D^{(j)})/\beta_0$. Here, $\gamma^{(1)}(O_D^{(j)})$ is the effective leading-order anomalous dimension of the $D$-dimensional OPE operator $O_D^{(j)}$, and $\beta_0$ is the one-loop $\beta$-function coefficient\footnote{We have $\beta_0 = (11 - 2 N_f/3)/4$; we use $N_f=3$, so $\beta_0=9/4$; cf.~also Eq.~(\ref{RGE}).} of the coupling $a(Q^2)$ [cf.~also the later discussion in Sec.~\ref{sec:renmod} around Eqs.~(\ref{Bdpk0})-(\ref{Btd5P})]. This means that the general form of the $D$-dimensional contribution to the OPE of the Adler function has the form
\bea
    {\cal D}_{{\rm th}, D=2 p}(Q^2)_{\rm V+A} &=& 2 p \pi^2 \frac{1}{(Q^2)^p} \sum_{j} \langle O_{2 p}^{(j)} \rangle_{\rm V+A} \; a(Q^2)^{k_D^{(j)}} \left( 1 + {\cal O}(a) \right)
\nonumber\\
    &=&  2 p \pi^2 \frac{1}{(Q^2)^p} \sum_{j} \langle O_{2 p}^{(j)} \rangle_{\rm V+A} \; a(Q^2)^{\gamma^{(1)}(O_D^{(j)})/\beta_0} \left( 1 + {\cal O}(a) \right),
\label{D2p} \eea
where the relative corrections ${\cal O}(a)$ are usually neglected.
For $D=4$ the situation is simpler, $D=4$ operator is predominantly gluon condensate, and has $\gamma^{(1)}(O_4)=0$. For $D=6$ we have nine operators. If we approximate the $D=6$ contribution by the first four operators, we notice that they have, among them, approximately just two different effective leading-order anomalous dimension values, $\gamma^{(1)}(O_6^{(1)})/\beta_0 \approx 0.222$ and $\gamma^{(1)}(O_6^{(2)})/\beta_0 \approx 0.625$, cf.~Eq.~(\ref{dD6}). We stress that in our previous work \cite{EPJ22,Trento} we applied for the effective leading-order anomalous dimensions (of the $D=4$ and $D=6$ operators) the large-$\beta_0$ approximation, which gives for $D=4$ unchanged result ($\gamma^{(1)}(O_4)=0$), but for $D=6$ gives significantly different values $\gamma^{(1)}(O_6^{(1)})/\beta_0 = -1$ and $\gamma^{(2)}(O_6^{(1)})/\beta_0 = 0$.
The anomalous dimensions of $D > 6$ operators are not known, and thus we will approximate those anomalous dimensions to be zero.  
We will thus use for the OPE expansion of the Adler function, improved with respect to the expansion (\ref{DOPE1}), the following truncated approximate form: 
\be
{\cal D}_{\rm th}(Q^2)_{\rm V+A} = d(Q^2)_{D=0} + 1 + 4 \pi^2 \frac{\langle O_{4} \rangle_{\rm V+A} }{ (Q^2)^2} + \frac{6 \pi^2}{(Q^2)^3} \sum_{j=1}^2 \langle O^{(j)}_{6} \rangle_{\rm V+A}  \; a(Q^2)^{k^{(j)}} + 2 \pi^2 \sum_{p =4}^{p_{\rm max}} \frac{p \langle O_{2 p} \rangle_{\rm V+A} }{ (Q^2)^p},
\label{DOPE}
\ee
which is truncated at a dimension $D = 2 p_{\rm max}$. At first we can consider that this OPE has two different condensates at $D = 6$, with the aforementioned approximate values of the anomalous dimensions $k^{(1)}=\gamma^{(1)}(O_6^{(1)})/\beta_0 \approx 0.222$ and   $k^{(2)}=\gamma^{(1)}(O_6^{(2)})/\beta_0 \approx 0.625$  \cite{BHJ}.\footnote{It can be checked that the term $6 \pi^2 a(Q^2)^{k^{(j)}} \langle O_6^{(j)} \rangle/(Q^2)^3$ corresponds in the polarisation function $\Pi(Q^2)$ [cf.~Eq.~(\ref{Ddef})] to the term $a(Q^2)^{k^{(j)}} \langle O_6^{(j)} \rangle/(Q^2)^3$ up to corrections ${\cal O}(a)$.}

Consideration of the V-channel gives a very similar OPE expansion
\be
{\cal D}_{\rm th}(Q^2)_{\rm V} = d(Q^2)_{D=0} + 1 + 4 \pi^2 \frac{2 \langle O_{4} \rangle_{\rm V} }{ (Q^2)^2} + \frac{6 \pi^2}{(Q^2)^3} \sum_{j=1}^2 2 \langle O^{(j)}_{6} \rangle_{\rm V} \;  a(Q^2)^{k^{(j)}} + 2 \pi^2 \sum_{p =4}^{p_{\rm max}} \frac{2 p \langle O_{2 p} \rangle_{\rm V} }{ (Q^2)^p},
\label{DOPEV}
\ee
i.e., the $D=0$ part is the same, and the condensates $\langle O_{2 p} \rangle_{\rm V+A}$ get replaced by $2 \langle O_{2 p} \rangle_{\rm V}$. We adhere to the notational convention $\langle O_{2 p} \rangle_{\rm V+A} = \langle O_{2 p} \rangle_{\rm V}+ \langle O_{2 p} \rangle_{\rm A}$, and ${\cal D}_{V+A} =(1/2)( {\cal D}_{V}+ {\cal D}_{A})$. Concerning the $D=6$ terms, in the V-channel we have an additional condensate (which violates chiral symmetry) with anomalous dimension $k^{(0)}=\gamma^{(1)}(O_6^{(0)})/\beta_0 = 0.111$ \cite{BHJ}.\footnote{There is also another chiral symmetry violating operator, but its anomalous dimension is significantly higher, $\gamma^{(1)}(O_6)/\beta_0=1.111$.} However, since this value is very close to the value of $k^{(1)} \approx 0.222$, we will approximate the sum of the two contributions of the corresponding condensates to be merged into one contribution with anomalous dimension $k \approx k^{(1)}$ ($=0.222$).

Later in Sec.~\ref{sec:fitBL}, it will turn out that the fits, in both (V+A)-channel and V-channel, lead to strong cancellations between the two $D=6$ terms and thus to numerical instabilities, especially when terms with $D \geq 8$ are included in the analysis. Consequently, we will then proceed with the OPE form (\ref{DOPE}) and (\ref{DOPEV}) with only one $D=6$ term, namely the one which is formally leading in power of $a$  ($\sim a^{k^{(1)}} = a^{0.222}$).

In the expressions (\ref{DOPE}) and (\ref{DOPEV}), the relative corrections ${\cal O}(a)$ were not included.

\begin{figure}[htb] 
\centering\includegraphics[width=70mm]{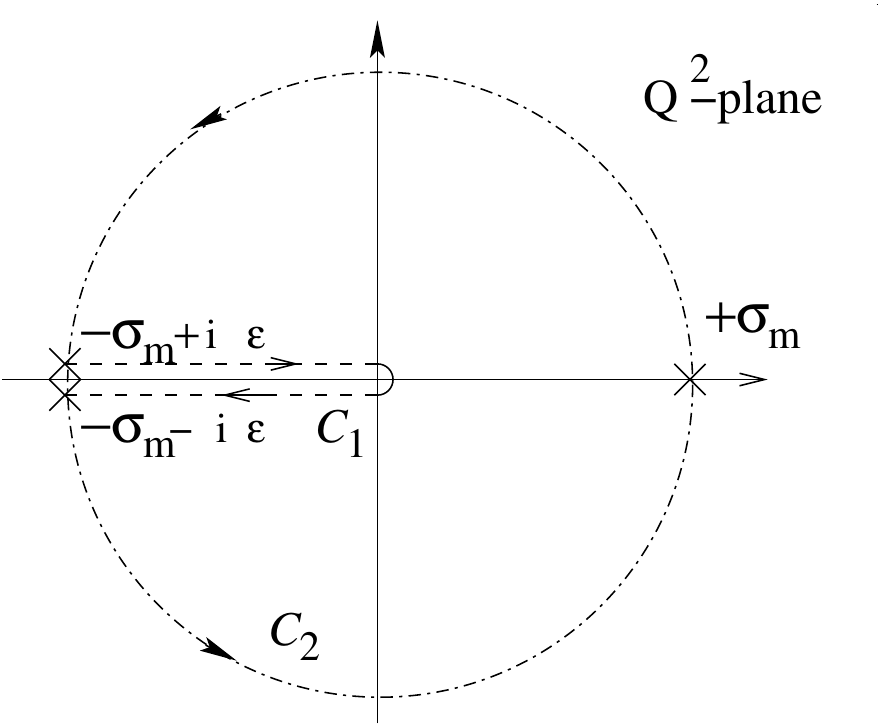}
\caption{\footnotesize The closed path $C_1+C_2$ of integration for $\oint d Q^2 g(Q^2) \Pi(Q^2)$. The circle $C_2$ has the radius $|Q^2|=\sigma_{\rm max}$ ($\equiv \sm$) ($\leq m_{\tau}^2$). It is understood that in the sector $C_1$ the limit $\varepsilon \to +0$ is taken.}
\label{Figcont2}
\end{figure}

\begin{figure}[htb] 
\begin{minipage}[b]{.49\linewidth}
  \centering\includegraphics[width=85mm]{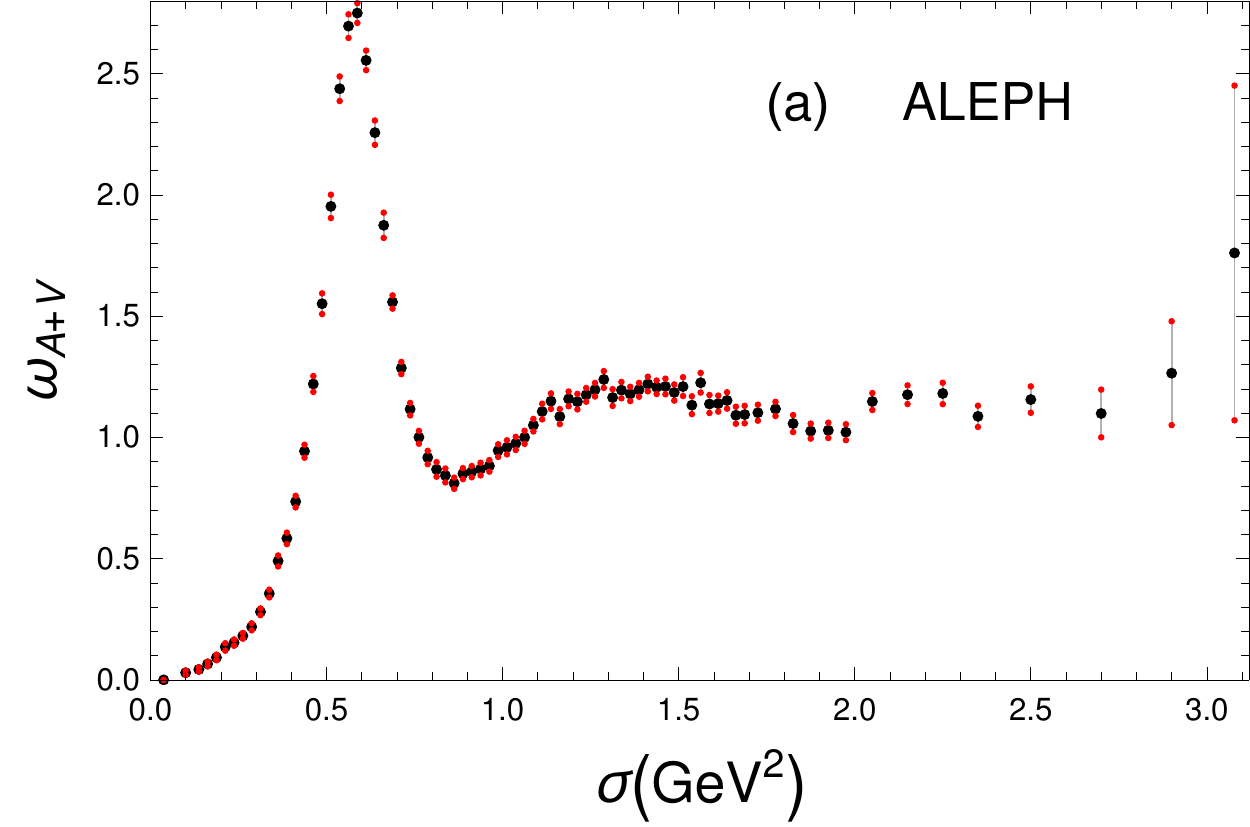}
  \end{minipage} 
\begin{minipage}[b]{.49\linewidth}
  \centering\includegraphics[width=85mm]{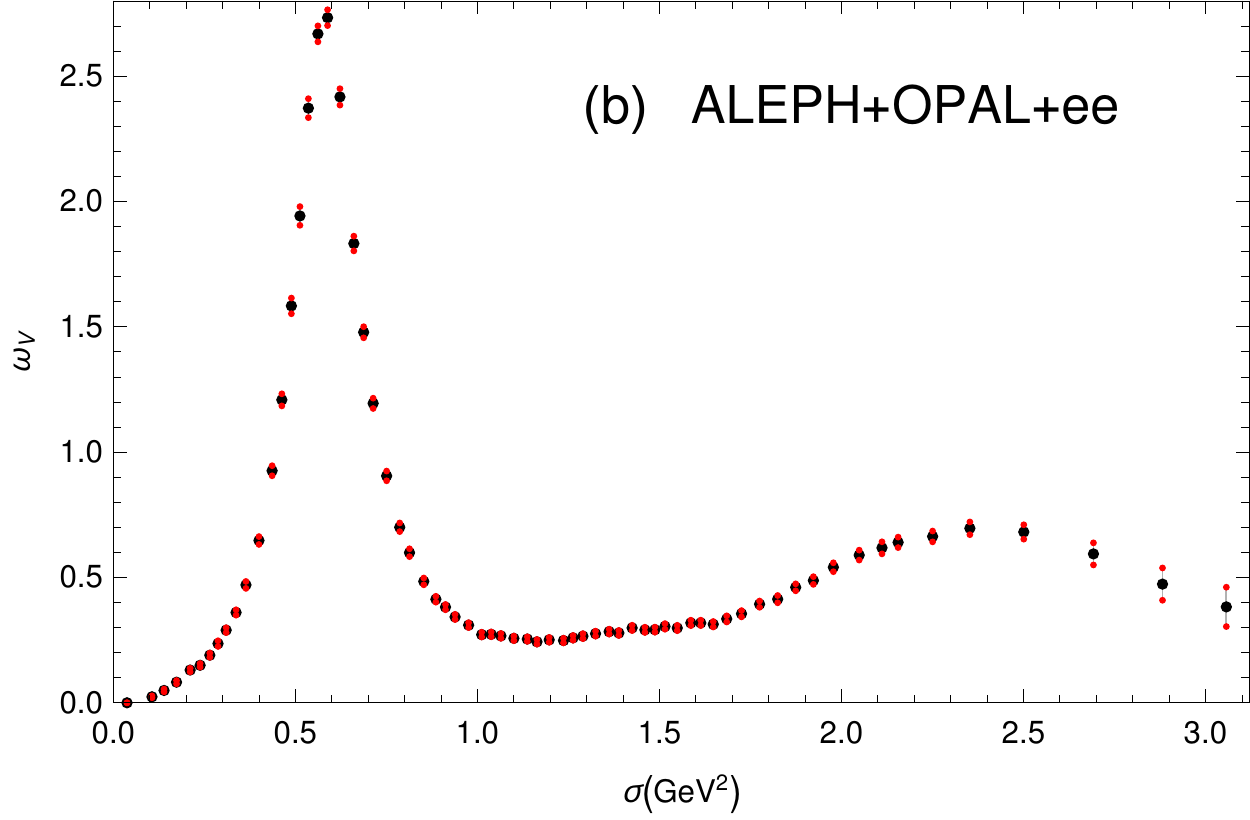}
\end{minipage}
\vspace{-0.2cm}
\caption{\footnotesize  (coloured online) (a) The spectral function $\omega(\sigma)_{\rm V+A}$ for the full (V+A)-channel, measured by ALEPH Collaboration \cite{ALEPH2,DDHMZ,ALEPHfin,ALEPHwww}. The pion peak contribution $2 \pi^2 f^2_{\pi} \delta(\sigma - m^2_{\pi})$ ($f_{\pi}=0.1305$ GeV) is not included in the Figure, and has to be added. We do not include the last two bins in our analysis, due to large uncertainties; therefore, $\sm =2.80 \ {\rm GeV}^2$ in the sum rules. (b) The spectral function  $\omega(\sigma)_{\rm V}$ for the V-channel \cite{Boito:2020xli,Perisetal}, based on a combination of ALEPH, OPAL and electroproduction data.}
\label{FigOmega}
\end{figure}

The Adler function ${\cal D}(Q^2)$ is a spacelike QCD observable, and is thus a holomorphic (analytic) function in the complex $Q^2$-plane with the exception of (a part of) the negative semiaxis $Q^2 < 0$, according to the general principles of Quantum Field Theory \cite{Bogolyubov:1959bfo,Oehme:1994pv}. When the integral $\oint dQ^2 g(Q^2) \Pi(Q^2;\mu^2)$ is considered, where $g(Q^2)$ is any holomorphic function and the integration is along the closed path of Fig.~\ref{Figcont2},
the following sum rule associated with $g$-function is obtained:
\bes
\label{Cauchy}
\bea
\oint_{C_1+C_2} d Q^2 g(Q^2) \Pi(Q^2) & = & 0  
\label{Cauchya} \\
\Rightarrow \;\; \int_0^{\sm} d \sigma g(-\sigma) \omega_{\rm exp}(\sigma)  &=&
-i \pi  \oint_{|Q^2|=\sm}
d Q^2 g(Q^2) \Pi_{\rm th}(Q^2) .
\label{Cauchyb} \eea \ees
The integration on the right-hand side of Eq.~(\ref{Cauchyb}) is counterclockwise. On the left-hand side, $\omega(\sigma)$ is the discontinuity (spectral) function of the (V+A)-channel polarisation function
\be
\omega(\sigma) \equiv 2 \pi \; {\rm Im} \ \Pi(Q^2=-\sigma - i \epsilon) \ .
\label{om1}
\ee
This quantity $\omega(\sigma)$ was measured in the semihadronic strangeless decays of the $\tau$ lepton, by OPAL \cite{OPAL,PerisPC1} and by ALEPH Collaboration \cite{ALEPH2,DDHMZ,ALEPHfin,ALEPHwww}. In our analysis we will use: the ALEPH data for the channel V+A, cf.~Fig.~\ref{FigOmega}(a); and the V-channel data based on combined ALEPH and OPAL data supplemented by electroproduction data \cite{Boito:2020xli,Perisetal}, cf.~Fig.~\ref{FigOmega}(b).
In the sum rule (\ref{Cauchyb}), the theoretical polarisation function is replaced by the Adler function (\ref{Ddef}) when the integration by parts is performed
\be
\int_0^{\sm} d \sigma g(-\sigma) \omega_{\rm exp}(\sigma)  =
\frac{1}{2 \pi}   \int_{-\pi}^{\pi}
d \phi \; {\cal D}_{\rm th}(\sm e^{i \phi}) G(\sm e^{i \phi}) .
\label{sr}
\ee 
The total Adler function ${\cal D}_{\rm th}(Q^2)$  is given by the OPE expansion (\ref{DOPE}); the function $G$ is an integral of the function $g$:
\be
G(Q^2)= \int_{-\sm}^{Q^2} d Q^{'2} g(Q^{'2}).
\label{GQ2}
\ee
The Adler function ${\cal D}(Q^2)$ is a spacelike physical quantity, i.e., it is a holomorphic function of $Q^2$ in the complex $Q^2$-plane with the exception of the negative semiaxis $Q^2<0$, as mentioned earlier. However, the sum rule (\ref{sr}) is a function of $Q^2=-\sigma <0$ ($-\sm \leq -\sigma <0$), i.e., it is a timelike physical quantity [as is also the spectral function $\omega(\sigma)$]. Other examples of the timelike physical quantities that are integrals involving the Adler function are \cite{Nesterenko:2016pmx}: (a) the ratio $R(\sigma)$ for $e^+ e^- \to$ hadrons production \cite{AKR,ANR} (where $-Q^2 = \sigma >0$ can take any value); (b) the leading order contribution of the hadronic vacuum polarisation to the muon anomalous magnetic moment $(g_{\mu}-2)$ \cite{amurev,amuZoltan} (where the dominant contributions come from $-Q^2 = \sigma \sim m^2_{\mu} \sim 0.01 \ {\rm GeV}^2$) \cite{NestJPG42,amuO}.

We recall that  the theoretical OPE expression (\ref{DOPE}) for the full Adler function ${\cal D}(Q^2)$ includes the $D=0$ contribution $d(Q^2)_{D=0}$, which can be regarded as the perturbative QCD part of ${\cal D}(Q^2)$. The first four perturbation coefficients $d_n$ in $d(Q^2)_{D=0} = \sum d_n a(Q^2)^{n+1}$  (i.e., $d_n$ for $n=0,1,2,3$; $d_0=1$) are known exactly \cite{d1,d2,BCK}, and there are estimates for $d_4$. In the next Section \ref{sec:renmod} we will present a renormalon-motivated model, related to the OPE (\ref{DOPE}), that generates the values of the other higher order coefficients $d_n$ (in fact, their estimates). Therefore, in the sum rules (\ref{sr}), the theoretical parameters to fit will be: the value of the QCD coupling $a(\sm^2)$ [$\leftrightarrow \alpha_s(M_Z^2)$] and the values of the condensates appearing in the OPEs (\ref{DOPE}) and (\ref{DOPEV}). When considering the (V+A)-channel, the spectral function $\omega_{\rm exp}(\sigma)$ on the left-hand side of Eq.~(\ref{sr}) will be $\omega_{\rm V+A}$ measured by ALEPH \cite{ALEPH2,DDHMZ,ALEPHfin,ALEPHwww} (cf.~Fig.~\ref{FigOmega}(a) for  $\omega_{\rm V+A}$). On the other hand, when considering the V-channel, $\omega_{\rm exp}(\sigma)$ will be $2 \omega(\sigma)_{\rm V}$ (cf.~Fig.~\ref{FigOmega}(b) for $\omega_{\rm V}$), according to our conventions [cf.~the OPEs Eqs.~(\ref{DOPEV}) and (\ref{DOPE})], where $\omega(\sigma)_{\rm V}$ \cite{Boito:2020xli,Perisetal} combines the V-channel data of ALEPH and OPAL and electroproduction data.

In the sum rules (\ref{sr})-(\ref{GQ2}), we will consider the weight functions $g(Q^2)$ of the double-pinched Borel-Laplace transforms $B(M^2)$ \cite{EPJ21,EPJ22}, which are functions of a complex squared energy parameter $M^2$
\bes
\label{BL}
\bea
g_{M^2}(Q^2) &=&  \left( 1 + \frac{Q^2}{\sm} \right)^2  \frac{1}{M^2} \exp \left( \frac{Q^2}{M^2} \right) \qquad \Rightarrow
\label{gM2} \\
G_{M^2}(Q^2) & = &  \left\lbrace  \left[ \left( 1 + \frac{Q^2}{\sigma_m} \right)^2 - 2\frac{M^2}{\sigma_m} \left(1 + \frac{Q^2}{\sigma_m}\right) + 2 \left(\frac{M^2}{\sigma_m} \right)^2 \right] \exp\left(\frac{Q^2}{M^2}\right) - 2 \left( \frac{M^2}{\sigma_m} \right)^2 \exp \left( -\frac{\sigma_m}{M^2}\right) \right\rbrace.
\label{GM2}
\eea \ees
Here, ``double-pinched'' means that $g(Q^2)$ has double zero at the timelike point $Q^2=-\sm$, and this suppresses strongly the duality violation effects, cf.~e.g.~Ref.~\cite{Pich,PichDV}. The weight function $G(Q^2)$ has then triple zero at $Q^2=-\sm$.
Taking into account the OPE (\ref{DOPE}), the theoretical expression for the Borel-Laplace sum rule $B_{\rm th}(M^2)$ is then
\bea
\lefteqn{
B_{\rm th}(M^2; \sm) =   \frac{1}{2 \pi} \int_{-\pi}^{+\pi} d \phi \;
G_{M^2} \left(\sm e^{i \phi} \right) {\cal D}_{\rm th} \left( \sm e^{i \phi} \right)
}
  \nonumber\\ 
  &=&
  \left[ \left( 1 - 2 \frac{M^2}{\sm} \right) + 2 \left( \frac{M^2}{\sm} \right)^2 \left(1 - \exp \left( - \frac{\sm}{M^2} \right) \right) \right]  
  \nonumber\\ &&
+ \frac{1}{2 \pi}  \int_{-\pi}^{+\pi} d \phi \left\{ \left[ \left(1 + e^{i \phi} \right)^2 - 2 \frac{M^2}{\sm}  \left(1 + e^{i \phi} \right) +  2 \left( \frac{M^2}{\sm} \right)^2 \right] \exp \left( \frac{\sm}{M^2} e^{i \phi} \right) -  2 \left( \frac{M^2}{\sm} \right)^2 \exp \left( - \frac{\sm}{M^2} \right) \right\} d \left( \sm e^{i \phi} \right)_{D=0}
  \nonumber\\ &&
 + \sum_{D=4,8,\ldots} B_{\rm th}(M^2; \sm)_{D} + B_{\rm th}(M^2; \sm)_{D=6},
\label{Bth}
\eea
where we have for $D >0$ and $D \not= 6$ ($D = 2 p$, $p=2,4,5,\ldots$)
\be
B_{\rm th}(M^2; \sm)_{D=2 p} = \frac{2 \pi^2 \langle O_{2 p} \rangle_{\rm V+A}}{ (p-1)! (M^2)^p } \left[ 1 + 2 (p-1) \frac{M^2}{\sm} + (p-1)(p-2) \left( \frac{M^2}{\sm} \right)^2 \right],
\label{BD}
\ee
and we evaluate the $D=6$ contribution by using there the coupling $a(\sm e^{i \phi} )$ running according to the one-loop RGE along the contour (since only the leading power of the coupling is known there)
\bea
B_{\rm th}(M^2; \sm)_{D=6} &=& \frac{1}{2 \pi} \frac{6 \pi^2}{\sm^3} \sum_{j=1}^2 \langle O^{(j)}_{6} \rangle_{\rm V+A} \int_{-\pi}^{+\pi} d \phi \; e^{- i 3 \phi} G_{M^2}(\sm e^{i \phi}) \left( \frac{a(\sm)}{1 + i \beta_0 \phi a(\sm)} \right)^{k^{(j)}}.
\label{BD6} \eea
We will use the above Borel-Laplace sum rules to extract the values of the coupling and of the condensates. Subsequently, we will evaluate in addition also the double-pinched momenta $a^{(2,n)}(\sm)$ (with $n=0,1$). The momenta $a^{(2,n)}(\sm)$ have the following weight functions $g^{(2,n)}(Q^2)$ ($n=0,1,\ldots$):
\bes
\label{FESR}
\bea
g^{(2,n)}(Q^2) &=& \left( \frac{n + 3}{n + 1} \right)\frac{1}{\sigma_m} \left( 1 + \frac{ Q^2}{\sigma_m} \right)^2 \sum_{k = 0}^{n} (k + 1)(-1)^k\left(\frac{Q^2}{\sigma_m} \right)^k \nonumber \\
 &=& \left( \frac{n + 3}{n + 1} \right)\frac{1}{\sigma_m} \left[ 1 - (n + 2) \left( - \frac{Q^2}{\sigma_m}\right)^{n + 1}  + (n + 1) \left(-\frac{Q^2}{\sigma_m} \right)^{n + 2}  \right] \; \Rightarrow 
\label{g2n}
\\
 G^{(2,n)}(Q^2) &=& \left( \frac{n + 3}{n + 1} \right)\frac{Q^2}{\sigma_m} \left[ 1 - \left(- \frac{Q^2}{\sigma_m} \right)^{n + 1}   \right] + \left[1 -  \left(-\frac{Q^2}{\sigma_m} \right)^{n + 3} \right].
\label{G2n}
\eea \ees
When we apply these weight functions to the sum rules (\ref{sr}), and subtract unity, we obtain the momenta $a^{(2,n)}$
\bes
\label{a2n}
\bea
a_{\rm exp}^{(2,n)}(\sigma_m) &=& \int_{0}^{\sigma_m} d\sigma \; g^{(2,n)}(-\sigma) \omega_{\rm exp}(\sigma) - 1,
\label{a2nexp} \\
 a_{\rm th}^{(2,n)}(\sigma_m) &=&  \frac{1}{2\pi} \int_{-\pi}^{\pi} d\phi \; G^{(2,n)}(\sigma_m e^{i\phi}) \left[ D_{\rm th}(\sigma_m e^{i\phi}) - 1 \right].
 \label{a2nth} \eea \ees
 Due to zero anomalous dimensions of the terms with $D \not=6$ in the OPE, it turns out that $a_{\rm th}^{(2,n)}(\sigma_m)$ depend, in addition to the $D=6$ condensates, on at most two OPE condensates
\bea
a_{\rm exp}^{(2,n)}(\sigma_m) &=& \frac{1}{2\pi} \int_{-\pi}^{\pi} d\phi \; G^{(2,n)}(\sigma_m e^{i\phi}) d(\sigma_m e^{i\phi})_{D=0}
\nonumber\\ &&
+ \frac{1}{2 \pi} \sum_{j=1}^2 \frac{6 \pi^2 \langle O_6^{(j)} \rangle_{\rm V+A}}{\sm^3} \int_{-\pi}^{\pi} d\phi \; e^{- i 3 \phi} G^{(2,n)}(\sigma_m e^{i\phi}) \left( \frac{a(\sm)}{1 + i \beta_0 \phi a(\sm)} \right)^{k^{(j)}}
\nonumber\\
&& + \left( \frac{n+3}{n+1} \right) 2 \pi^2 (-1)^n \left\{ (n+2) \frac{\langle O_{2 n + 4} \rangle_{\rm V+A}}{\sm^{n+2}} +  (n+1) \frac{\langle O_{2 n + 6} \rangle_{\rm V+A}}{\sm^{n+3}} \right\}.
\label{a2nthexp}
\eea
In this expression, the last two condensates enter for $n \geq 2$. For $a^{(2,0)}$ only the first of these two condensates enters ($\langle O_4 \rangle$), and for $a^{(2,1)}$ only the second ($\langle O_8 \rangle$). The contour integral involving $\langle O_6^{(j)} \rangle_{\rm V+A}$ enters for all $n$. When the V-channel is considered, the mentioned replacement $\langle O_D \rangle_{\rm V+A} \mapsto 2 \langle O_D \rangle_{\rm V}$ has to be made.

As in the case of the Borel-Laplace sum rules, Eq.~(\ref{BD6}), the coupling $a(\sm e^{i \phi})$ in the $D=6$ contributions will be taken as running according to the one-loop RGE along the contour. We note that $a^{(2,1)}(\sigma=m^2_{\tau})$ is the canonical QCD (and strangeless and massless) $\tau$-decay ratio $r_{\tau}$.

For the V-channel, the same formulas Eqs.~(\ref{Bth})-(\ref{a2nthexp}) apply, but with the mentioned replacements $\langle O_{2 p} \rangle_{\rm V+A} \mapsto 2 \langle O_{2 p} \rangle_{\rm V}$ wherever the condensates appear.

\section{Renormalon-motivated extension of the Adler function}
\label{sec:renmod}

Here we present construction of a renormalon-motivated extension of the Adler function, following the approach of \cite{renmod} (see also \cite{EPJ21}).\footnote{An extensive review of the topics on QCD renormalons until 1999, including many references, is given in \cite{ren}. Since then, various other works on renormalons have appeared, among them \cite{Maiezza,Cavalc,BoiOl,Pineda1,Pineda2}.} The expansion of the leading-twist ($D=0$) contribution to the Adler function in powers of $a \equiv \alpha_s/\pi$ has the form
\be
d(Q^2)_{D=0, {\rm pt}}= d_0 a(\kappa Q^2) + d_1(\kappa) \; a(\kappa Q^2)^2 + \ldots + d_n(\kappa) \; a(\kappa Q^2)^{n+1} + \ldots, \qquad (d_0=1),
\label{dpt}
\ee
where the exact values of the first four expansion coefficients ($d_0=1; d_1; d_2; d_3$) are known \cite{d1,d2,BCK}. We denoted $\kappa \equiv \mu^2/Q^2$ ($0 <\kappa \sim 1$), i.e., $\kappa$ is the dimensionless parameter for the renormalisation scale $\mu^2$. We then rearrange this power expansion (\ref{dpt}) into the expansion in terms of the logarithmic derivatives $\ta_{n+1}$
\be
\ta_{n+1}(Q^{'2}) \equiv \frac{(-1)^n}{n! \beta_0^n} \left( \frac{d}{d \ln Q^{'2}} \right)^n a(Q^{'2}) \qquad (n=0,1,2,\ldots).
\label{tan} \ee
The reorganised expansion (logarithmic perturbation series 'lpt') is
\be
d(Q^2)_{D=0, {\rm lpt}}= {\td}_0 a(\kappa Q^2) + {\td}_1(\kappa) \; {\ta}_2(\kappa Q^2) + \ldots + {\td}_n(\kappa) \; {\ta}_{n+1}(\kappa Q^2) + \ldots.
\label{dlpt} \ee
We work in the perturbative $\MSbar$ scheme \cite{MSbar}, whose RGE is known up to five-loops \cite{5lMSbarbeta}
\be
\frac{d a(\kappa Q^2)}{d \ln \kappa} = - \beta_0 a(\kappa Q^2)^2 -\beta_1 a(\kappa Q^2)^3 - \sum_{j=2}^4 {\beta}_j a(\kappa Q^2)^{j+2},
\label{RGE}
\ee
where $\beta_0$ and $\beta_1$ are universal (i.e., scheme independent), $\beta_0 = (11 - 2 N_f/3)/4$ ($=9/4$ for $N_f=3$) and $\beta_1=(1/16)(102 - 38 N_f/3)$, while $\beta_j$ for $j \geq 2$ have specific scheme dependent values (here: $\MSbar$ values). We use for the effective number of quark flavors $N_f=3$ throughout.   
Knowing the $\beta_j$ coefficients, the relations between the logarithmic derivatives $\ta_{n+1}$ and the powers $a^k$ ($k \geq n+1$) can be found, and they have the form
\bes
\label{tananantan}
\bea
\ta_{n+1}(Q^{'2}) &=& a(Q^{'2})^{n+1} + \sum_{m \geq 1} k_m(n+1) \; a(Q^{'2})^{n+1+m},
\label{tanan} \\
a(Q^{'2})^{n+1} &=&  \ta_{n+1}(Q^{'2}) + \sum_{m \geq 1} \tk_m(n+1) \; \ta_{n+1+m}(Q^{'2}).
\label{antan} \eea \ees
Further, the corresponding expansion coefficients $\td_n$ and $d_k$ are related analogously
\bes
\label{dnvstdn}
\bea
\td_n(\kappa) &=& d_n(\kappa) + \sum_{s=1}^{n-1} \tk_s(n+1-s) \; d_{n-s}(\kappa) \qquad (\td_0=d_0=1),
\label{tdndk} \\
d_n(\kappa) &=& \td_n(\kappa) + \sum_{s=1}^{n-1} k_s(n+1-s) \; \td_{n-s}(\kappa) \quad
(n=0,1,2, \ldots).
\label{dntdk} \eea \ees
Here, the coefficients $\tk_s(n+1-s)$ and $k_s(n+1-s)$ are specific ($\kappa$-independent) expressions of the $\beta$-function coefficients $c_j \equiv \beta_j/\beta_0$, cf.~\cite{renmod}. For example, the first few coefficients $k_s$ have the following form:
\bes
\label{ks}
\bea
k_1(2)& = & c_1, \quad k_1(3) = \frac{5}{2} c_1, \quad k_2(2)= c_2,
\label{k22} \\
k_1(4) &=& \frac{13}{3} c_1, \quad k_2(3)=3 c_2 + \frac{3}{2} c_1^2,
\quad k_3(2)=c_3,
\label{k32}
\eea \ees
etc., where $c_j \equiv \beta_j/\beta_0$, and the coefficients $\beta_j$ appear in the RGE (\ref{RGE}). These coefficients are independent of the renormalisation scale parameter $\kappa$.

The expansion coefficients $d_n$, and the new expansion coefficients $\td_n$, allow us to construct the Borel transforms ${\cal B}[d](u)$ and ${\cal B}[\td](u)$
\bes
\label{Bdtdexp}
\bea
    {\cal B}[d](u; \kappa)_{\rm ser.} &\equiv& d_0 + \frac{d_1(\kappa)}{1! \beta_0} u + \ldots + \frac{d_n(\kappa)}{n! \beta_0^n} u^n + \ldots \ ,
    \; \Rightarrow
\label{Bdexp} \\
{\cal B}[\td](u; \kappa)_{\rm ser.} &\equiv& {\td}_0 + \frac{{\td}_1(\kappa)}{1! \beta_0} u + \ldots + \frac{{\td}_n(\kappa)}{n! \beta_0^n} u^n + \ldots.
\label{Btdexp} \eea \ees
The transform ${\cal B}[\td](u)$ is the Borel transform of an auxiliary quantity $\td(Q^2; \kappa)$ to the Adler function
\be
\td(Q^2; \kappa)_{\rm ser} \equiv a(\kappa Q^2) + \td_1(\kappa) a(\kappa Q^2)^2 + \ldots + \td_n(\kappa)  a(\kappa Q^2)^{n+1} + \ldots,
\label{tdpt} \ee
where this quantity has $\kappa$-dependence, in contrast to the Adler function Eq.~(\ref{dlpt}). Only in the one-loop approximation the auxiliary quantity $\td(Q^2; \kappa)$ and the Adler function $d(Q^2)_{D=0}$ coincide (then: $\ta_{n+1}=a^{n+1}$, ${\td}_n=d_n$).

First we discuss the singularity structure of the Borel transform  ${\cal B}[d](u)$ at $u >0$ (IR renormalons). Namely, the structure should be of such a form that the renormalon ambiguity, which appears upon integration in the inverse Borel transform, has the same $Q^2$-dependence as the corresponding OPE terms (of dimension $D > 0$); then it is possible to argue that the renormalon ambiguity of the $D=0$ term (obtained by the inverse Borel transform) can be cancelled by the corresponding $D>0$ contribution in the OPE. If the IR renormalon term with a pole (branching point) at $u=p > 0$ has the form
\be
{\cal B}[d](u)_{p, \kappa_p^{(j)}}^{(0)} = \frac{1}{(p- u)^{\kappa_p^{(j)} + p c_1/\beta_0}},
\label{Bdpk0} \ee     
then the corresponding renormalon ambiguity has the following $Q^2$-dependence:
\bes
\label{ImBdpk}
\bea
\delta d (Q^2)_{p,\kappa_p^{(j)}} & \sim & 
\frac{1}{\beta_0} {\rm Im} \int_{+ i \varepsilon}^{+\infty + i \varepsilon} d u \exp \left( - \frac{u}{\beta_0 a(Q^2)} \right)
\frac{1}{(p - u)^{\kappa_p^{(j)} + p c_1/\beta_0}}
\label{ImBdpka} \\
& \sim & \frac{1}{(Q^2)^p} a(Q^2)^{1 -\kappa_p^{(j)}}  \left[ 1 + {\cal O}(a) \right].
\label{ImBdpkb}
\eea \ees
This is the same $Q^2$-dependence as in the $D= 2 p$ dimensional OPE contribution Eq.~(\ref{D2p}) to the Adler function, once we identify
\be
p = D/2, \quad \kappa_p^{(j)} =1 - k_D^{(j)} = 1 - \gamma^{(1)}(O_D^{(j)})/\beta_0.
\label{ImBdpkD2p} \ee
This means that the IR renormalon expressions (\ref{Bdpk0}) in the Borel transform ${\cal B}[d](u)$ of the $D=0$ part of the Adler function, $d(Q^2)_{D=0}$ [cf.~Eq.~(\ref{DOPE})] reflect one-by-one the terms in the $D = 2 p >0$ OPE part of the Adler function Eq.~(\ref{D2p}). We also note that the additional term $p c_1/\beta_0$ ($=p \beta_1/\beta_0^2$) in the IR renormalon index in Eq.~(\ref{Bdpk0}) is a reflection of the effects of the RGE running of $a(Q^2)$ beyond the one-loop (i.e., beyond $\beta_0$). The above reasoning refers to those OPE terms Eq.~(\ref{D2p}) which have the condensates $\langle O_{2 p} \rangle$ of nonchiral origin, i.e., not connected with the generation of nonzero masses. In that context, we note that $d(Q^2)_{D=0}$ is a purely massless quantity.

We now turn to the corresponding IR renormalon terms in the Borel transform ${\cal B}[\td](u)$ of the auxiliary quantity $\td(Q^2; \kappa)$. We point out that the transform ${\cal B}[\td](u)$ contains all the information about the Adler function coefficients $\td_n$ (and thus $d_n$). Nonetheless, in contrast to  ${\cal B}[d](u)$, it has the simple one-loop type renormalisation scale dependence
\be
\frac{d}{d \ln \kappa} {\td}_n(\kappa) = n \beta_0 {\td}_{n-1}(\kappa) \quad \Rightarrow \quad {\cal B}[\td](u; \kappa) = \kappa^u {\cal B}[\td](u).
\label{tdnkap} \ee
This suggests that the renormalon structure of ${\cal B}[\td](u)$ has no explicit effects coming from the beyond one-loop RGE running of the coupling, i.e., no terms $\propto \beta_1$ in the power indices of the singularities 
when compared to the corresponding IR renormalon terms Eq.~(\ref{Bdpk0})
\be
{\cal B}[\td](u)_{p, \kappa_p^{(j)}}^{(0)} = \frac{1}{(p- u)^{\kappa_p^{(j)}}}.
\label{Btdpk0} \ee
The suggested behaviour Eq.~(\ref{Btdpk0}) turns out to be correct, as shown numerically in Ref.~\cite{renmod} for various cases of integer-valued $\kappa_p^{(j)}$. It can be shown numerically that this holds even when $\kappa_p^{(j)}$ are not integer-valued. Specifically, following the numerical approach of \cite{renmod}, we deduce from the expression (\ref{Btdpk0}) the corresponding expansion coefficients $\td_n$, and then $d_n$ via the relations (\ref{dntdk}), and the subsequent consideration of the numerical behaviour of the obtained $d_n$ for high $n$ gives us the following Borel transform ${\cal B}[d](u)_{p, \kappa_p^{(j)}}$:
\bes
\label{BtdpkvsBdpk}
\bea
{\cal B}[\td](u)_{p, \kappa_p^{(j)}} &=& \frac{\pi \td^{\rm IR}_{p,j}}{(p- u)^{\kappa_p^{(j)}}}
\; \Rightarrow
\label{Btdpk}
\\
{\cal B}[d](u)_{p, \kappa_p^{(j)}} &=& \frac{\pi d^{\rm IR}_{p,j}}{(p- u)^{\kappa_p^{(j)}+ p c_1/\beta_0}} {\Big \{} 1 + \frac{(b_1^{(2 p)} + {\cal C}^{(2 p)}_{1,j})}{\beta_0 (\kappa_p^{(j)}+ p c_1/\beta_0 -1)} (p - u) +
\nonumber\\
&& + \frac{(b_2^{(2 p)} + b_1^{(2 p)} {\cal C}^{(2 p)}_{1,j} + {\cal C}^{(2 p)}_{2,j})}{\beta_0^2  (\kappa_p^{(j)}+ p c_1/\beta_0 -1) (\kappa_p^{(j)}+ p c_1/\beta_0 -2)} (p - u)^2 + \cdots {\Big \}},
\label{Bdpk} \eea \ees
where the coefficients $b_j^{(2 p)}$ are those appearing in the expression for the power $1/(Q^2)^p$ in terms of $a(Q^2)^2$ ($\equiv a_Q$)
\be
\frac{1}{(Q^2)^p} = {\rm const} \exp \left( - \frac{p}{\beta_0 a_Q} \right) a_Q^{- p c_1/\beta_0} \left[ 1 + b_1^{(2 p)} a_Q + b_2^{(2 p)} a_Q^2 + {\cal O}(a^3) \right],
\label{1Q2p} \ee
i.e., their explicit expressions are
\bes
\label{b1b2}
\bea
b_1^{(2 p)} & = & \frac{p}{\beta_0} (c_1^2 - c_2),
\label{b1} \\
b_2^{(2 p)} & = & \frac{1}{2} \left( b_1^{(2 p)} \right)^2 - \frac{p}{2 \beta_0} \left(c_1^3 - 2 c_1 c_2 + c_3 \right).
\label{b2} \eea \ees
The coefficients ${\cal C}^{(2 p)}_{n,j}$ in expansion (\ref{Bdpk}) reflect certain combinations of the higher loop Wilson coefficients and the subleading effects of the anomalous dimension of the operator ${\cal O}_{2 p}^{(j)}$ \cite{renmod}. Furthermore, the ratio $d^{\rm IR}_{p,j}/\td^{\rm IR}_{p,j}$ of residues in Eqs.~(\ref{BtdpkvsBdpk}) also have specific numerical values. In Table \ref{tabratC} we present the values of these quantities, for the $\MSbar$ scheme, for the terms of our interest, i.e., $p=2$ and $k_4=0$ ($\kappa_2=1$); and $p=3$ and $k_6^{(1)}=0.222$ ($\kappa_3^{(1)}=0.778$) and $k_6^{(2)}=0.625$ ($\kappa_3^{(2)}=0.375$).\footnote{As we will argue later, near the leading IR renormalon location $p=2$, it is reasonable to include a subleading singularity $\sim \ln(1 - u/2)$ in  ${\cal B}[\td](u)$, which corresponds in the Borel transform of the Adler function to the subleading singularity ${\cal B}[d](u) \sim 1/(2-u)^{2 c_1/\beta_0}$, cf.~\cite{renmod}. We include this case in Table \ref{tabratC}, using the notations of \cite{renmod}.}
We include in Table \ref{tabratC} also the case of the first UV renormalon. The behaviour of the UV renormalon there is very similar to the IR case, except that now we have $-p <0$ instead of $p$. Specifically, for the leading UV renormalon, at $u=-p=-1$, we will take the approximation $\kappa_{-1}=2$ which is the dominant index in the large-$\beta_0$ (LB) approximation  \cite{LB1,LB2,ren}. For UV renormalons, we have relations analogous to Eqs.~(\ref{BtdpkvsBdpk}), and for convenience we present these relations in Appendix \ref{appUV1}.
\begin{table}
  \caption{The numerically extracted values of the coefficients  ${\cal C}^{(D)}_{1,j}$ and  ${\cal C}^{(D)}_{2,j}$, and the ratios  $d_{p,k}^{\rm X}/\td_{p,k}^{\rm X}$, for X=IR with $p=2,3$ ($D=2 p$), and X=UV and $-p=-1$ ($D=-2  p$) in the 5-loop $\MSbar$ renormalisation scheme. In the notation, $\pm p$ means that $+p$ is taken for IR, and $-p$ for UV cases. The cases $X=IR$ with $p=2$ and $X=UV$ with $-p=-1$ are those given in Table II of Ref.~\cite{renmod}. The case $p=2$ and $j=0$ is the case with ${\cal B}[\td](u)= \pi \td^{\rm IR}_{2,0}  (-1) \ln (1 - u/2)$, cf.~\cite{renmod}.}
\label{tabratC}
\begin{ruledtabular}
\begin{tabular}{lccc|c|cc}
  type (X) & $\pm p$ & $j$ & $\kappa_{\pm p}^{(j)}$ & $d_{p,j}^{\rm X}/\td_{p,j}^{\rm X}$ &  ${\cal C}^{(D)}_{1,j}$ & ${\cal C}^{(D)}_{2,j}$
\\
\hline 
X=IR & $p=2$ & $j=1$ & $1$ & $(+1.7995 \pm 0.0001)$ & $(-0.03 \pm 0.02)$ & $(+1.7 \pm 0.3)$
\\
X=IR & $p=2$ & $j=0$ & $0$ & $(+1.155 \pm 0.005)$ & $(+7.7 \pm 0.4)$ & $\cdots$
\\
\hline
X=IR & $p=3$ & $j=1$ & $0.778$ & $(+6.078 \pm 0.031)$ & $(+1.55 \pm 0.27)$ & $(+3.1 \pm 1.7)$
\\
X=IR & $p=3$ & $j=2$ & $0.375$ & $(+2.177 \pm 0.039)$ & $(+3.20 \pm 1.00)$ & $(+29.6 \pm 6.2)$
\\
\hline
X=UV & $-p=-1$ & $j=2$ & $2$ & $(+1.056 \pm 0.014)$ & $(-10.1 \pm 2.1)$ & $(-83. \pm 8.)$
\end{tabular}
\end{ruledtabular}
\end{table}
In Appendix \ref{appSL} we also present, for completeness, the analogous results for the subleading cases, i.e., when the power index $\kappa_{\pm p}^{(j)}$ is decreased by one or two units; those results are needed when we change the renormalisation scale parameter $\kappa \equiv \mu^2/Q^2$ from $\kappa=1$ to $\kappa \not=1$ (later on we applied $\kappa_{\rm max}=2$ and $\kappa_{\rm min}=2/3$ cases as variation of $\kappa$ in our numerical analysis).  

Based on these considerations, we make for ${\cal B}[\td](u)$ (with $\kappa=1$) in the $\MSbar$ scheme the ansatz which reflects the corresponding discussed singularities at $u=2, 3$ (IR renormalons, $p=2, 3$) and at $u=-1$ (UV renormalon, $-p=-1$) 
\bea
{\cal B}[\td](u) & = & \exp \left( \tK u \right) \pi {\Big \{}
\td_{2,1}^{\rm IR} \left[ \frac{1}{(2-u)} + \tal (-1) \ln \left( 1 - \frac{u}{2} \right) \right] + \frac{ \td_{3,1}^{\rm IR} }{(3 - u)^{\kappa^{(1)}}} + \frac{ \td_{3,2}^{\rm IR} }{(3 - u)^{\kappa^{(2)}}} + \frac{ \td_{1,2}^{\rm UV} }{(1 + u)^2} {\Big \}},
\label{Btd5P}
\eea
where in the $u=3$ renormalon terms we have noninteger power indices $\kappa^{(j)}= 1-\gamma^{(1)}(O_6^{(j)})/\beta_0$ ($=1 - k^{(j)}$), where $\kappa^{(1)} \approx 0.778$ and $\kappa^{(2)} \approx 0.375$, cf.~also the discussion after Eqs.~(\ref{dD6}) and (\ref{DOPE}).\footnote{In \cite{EPJ21,EPJ22} we used, in the ansatz for ${\cal B}[\td](u)$, at the $u=3$ renormalon terms the power indices in the large-$\beta_0$ approximation: $\kappa^{(1)}=2$ and $\kappa^{(2)}=1$.} For simplicity, in these indices we omit the subscript $D=6$, cf.~also Eqs.~(\ref{dD6}) and (\ref{DOPE}).
Further, the value of the parameter $\tal$, in the considered $\MSbar$ scheme, is $\tal =-0.255$ \cite{renmod}. Namely, the value of $\tal$ is determined by the knowledge of the subleading coefficient ${\hat c}^{(D=4)}_1$, which is a combination of the leading and the subleading-order anomalous dimension of the $D=4$ (gluon) operator and of the corresponding Wilson coefficient \cite{CPS,ren}. The other five parameters, i.e., the scaling parameter $\tK$ and the four residues ($\td_{2,1}^{\rm IR}, \td_{3,2}^{\rm IR}, \td_{3,1}^{\rm IR}, \td_{1,2}^{\rm UV}$) are determined by the knowledge of the first five coefficients $\td_n$, i.e, thus by the knowledge of $d_n$ ($n=0,1,2,3,4$), cf.~Eqs.~(\ref{dpt})-(\ref{dlpt}), (\ref{dnvstdn}), (\ref{Bdtdexp}). While for $n \leq 3$ the values of the coefficients $d_n$ of the Adler function are exactly known \cite{d1,d2,BCK}, the value of the coefficient $d_4$ is not (yet) exactly known.

Various estimates for the value of $d_4$ exist. The method of effective charge (ECH) \cite{ECH} gives an estimate $d_4=275$ \cite{KatStar,BCK}. An estimate based on Pad\'e approximants gives a similar estimate $d_4=277 \pm 51$ \cite{Boitoetal}; yet another similar estimate $d_4=283$ is obtained by extrapolation of the approximately geometric series behaviour of FO of $r_{\tau} ^{(D=0)}$ [$=a^{(2,1)}(m_{\tau}^2)_{D=0}$] \cite{BJ}. On the other hand, in \cite{renmod} a simple ansatz for ${\cal B}[\td](u)$ was used in the (lattice-related) MiniMOM renormalisation scheme, giving the estimate for $d_4$ which, transformed to the $\MSbar$ scheme, gave $d_4 \approx 338$. We will use,\footnote{A method based on a specific reorganisation of the perturbation series \cite{GKM}, which is a two-fold expansion in powers of the conformal anomaly $\beta(a)/a$ and of $a$, may give yet another interesting estimate for $d_4$.} as we did in \cite{EPJ22,Trento}, for the central value of $d_4$ the ECH estimate $d_4=275.$, and include the estimate $d_4=338.$ by variation
\be
d_4 = 275. \pm 63.
\label{d4est} \ee
The described procedure gives, at a chosen value of $d_4$, several solutions for the values of the parameters $\tK$ and the four mentioned residues. However, these solutions, with one exception, have large absolute values, and the values of the $p=3$ residues ($\td_{3,2}^{\rm IR}, \td_{3,1}^{\rm IR}$) are large and with opposite signs, indicating spurious effects of strong cancellations. One solution, though, has distinctly smaller absolute values of the parameters, and in Table \ref{renmodpar} we present the resulting values of this solution, for three values of $d_4$ parameter ($d_4=275.$, $275. -63.$, $275.+63.$).
\begin{table}
  \caption{The values of the parameters $\tK$ and of the renormalon residues $\td_{i,j}^{\rm X}$ (X=IR,UV) of the expression (\ref{Btd5P}), in the five-loop $\MSbar$ scheme, when the value of $d_4$ coefficient is taken as the central value or the border values of the estimate Eq.~(\ref{d4est}).}
\label{renmodpar}
\begin{ruledtabular}
\begin{tabular}{r|rrrrr}
  $d_4$ & $\tK$ & $\td_{2,1}^{\rm IR}$ & $\td_{3,1}^{\rm IR}$ &  $\td_{3,2}^{\rm IR}$ & $\td_{1,2}^{\rm UV}$ 
\\
\hline
275.                & 0.131144 & 1.09671 & 1.03356 & -0.992974 & -0.0120413
\\
275.  - 63.       & -0.409096 & 2.59142 & 2.49249 & -3.05914 & -0.0115282
\\
275.  + 63.       & 0.525533 & 0.98966 & -1.04184 & 0.420453 & -0.0117992
\end{tabular}
\end{ruledtabular}
\end{table}
Further, in Table \ref{tabtdndn} we present the corresponding expansion coefficients $\td_n$ and $d_n$ for these three cases, for $n \leq 10$.
\begin{table}
    \caption{The coefficients $\td_n$ and $d_n$ (in the $\MSbar$ scheme and with $\kappa=1$) for the three cases of $d_4$ Eq.~(\ref{d4est}), for $n \leq 10$.}
\label{tabtdndn}
\begin{ruledtabular}
\begin{tabular}{r|rr|rr|rr}
 $n$ & $d_4=275.$: $\td_n$ & $d_n$ & $d_4=212.$: $\td_n$ & $d_n$ & $d_4=338.$: $\td_n$ & $d_n$ 
\\
\hline
0       &  1       & 1           &  1       & 1  &  1       & 1   \\
1       &  1.63982 & 1.63982    &  1.63982 & 1.63982  &  1.63982 & 1.63982 \\
2       &  3.45578 & 6.37101    &  3.45578 & 6.37101  &  3.45578 & 6.37101\\
3       &  26.3849 & 49.0757   &  26.3849 & 49.0757 &  26.3849 & 49.0757\\
4       & -25.4180 & 275. &  -88.4180 & 212. & 37.5820 & 338. \\
5       &  1859.36 & 3206.48 & 2471.54 & 3099.99 & 1718.45 & 3784.23 \\
6       & -19035.2 & 16901.6 & -31100.9 & 8666.78 & -10053. & 29341.8 \\
7       &  421210. & 358634. & 628468. & 375914. & 319293. & 458189. \\
8       & $-7.80444 \times 10^{6}$ & 621177. & $-1.22286 \times 10^{7}$ & $-1.07315 \times 10^{6}$ & $-5.11629 \times 10^{6}$ & $3.09086 \times 10^{6}$ \\
9       & $1.82502 \times 10^{8}$ & $7.52194 \times 10^{7}$ & $2.82236 \times 10^{8}$ & $9.29524 \times 10^{7}$ &  $1.27983 \times 10^{8}$ & $8.66934 \times 10^{7}$ \\
10      & $-4.43137 \times 10^{9}$ & $-5.21168 \times 10^{8}$ & $-6.94764 \times 10^{9}$ & $-1.15783 \times 10^{9}$ & $-2.99926 \times 10^{9}$ & $1.43836 \times 10^{8}$
\end{tabular}
\end{ruledtabular}
\end{table}
In the expression (\ref{Btd5P}) for ${\cal B}[\td](u)$, we can expand the overall exponential factor $\exp(\tK u)$ in powers of $(p -u)$ ($p=2, 3$) for the IR terms, and powers of $(1+u)$ for the UV term.  As explained earlier in this Section, each such (singular) term in the obtained expression for ${\cal B}[\td](u)$ then leads to a corresponding series of singular terms in the Borel transform ${\cal B}[d](u)$  [cf.~Eqs.~(\ref{BtdpkvsBdpk}) and (\ref{BtdpkvsBdpkUV})]. Therefore, the expression (\ref{Btd5P}) of ${\cal B}[\td](u)$ leads to the following (renormalon) expansion of the Borel transform ${\cal B}[\td](u)$ (for $\kappa=1$; $d_4=275.$; $N_f=3$):
\bea
\lefteqn{
  \frac{1}{\pi} {\cal B}[d](u) \equiv  \frac{1}{\pi} \left[ d_0 + \frac{d_1}{1! \beta_0} u + \ldots + \frac{d_n}{n! \beta_0^n} u^n + \ldots \right]
  }
\nonumber\\
& = &
{\bigg \{} \frac{2.565390}{(2 - u)^{209/81}}
\left[ 1 - 0.499518 \; (2-u) -1.36093  \; (2-u)^2 + {\cal O}( (2-u)^3) \right] 
{\bigg \}}
\nonumber\\ &&
+ {\bigg \{}
 \frac{9.310304}{(3 - u)^{3.14837}} \left[ 1 - 0.039861 \; (3-u) - 0.415167  \; (3-u)^2 + {\cal O}( (3-u)^3) \right]
\nonumber\\ &&
+ \frac{(-3.203771)}{(3 - u)^{2.74537}} \left[ 1 +0.418500  \; (3-u)  + {\cal O}( (3-u)^2) \right]
{\bigg \}}
\nonumber\\ &&
+ {\bigg \{}
 \frac{(-0.0111528)}{(1+ u)^{98/81}} \left[ 1 + 20.7769   \; (1+u) +  103.042  \; (1 + u)^2  + {\cal O} ( (1+u)^3) \right] 
{\bigg \}}.
\nonumber\\
\label{Bd5P}
\eea
The expressions for the cases $d_4=275.-63.$ and $d_4=275.+63.$ are obtained analogously. Further, when the renormalisation scale parameter $\kappa$ is changed from $\kappa=1$ to $\kappa \not=1$ ($\kappa=2/3$; $\kappa=2$), then the expression ${\cal B}[\td](u; \kappa)$ is obtained by multiplication by the simple exponential factor $\exp( u \ln \kappa)$, cf.~Eq.~(\ref{tdnkap}), and the corresponding expansion for ${\cal B}[d](u; \kappa)$ is obtained in analogous way as ${\cal B}[d](u)$ in Eq.~(\ref{Bd5P}).

One may ask why our approach is based on an ansatz for the Borel transform ${\cal B}[\td]$ of the auxiliary quantity $\td(Q^2)$ Eq.~(\ref{tdpt}) and not on an ansatz of the Borel transform ${\cal B}[d]$ of the ($D=0$) Adler function $d(Q^2)_{D=0}$ itself Eq.~(\ref{dpt}). As explained in Ref.~\cite{renmod}, the knowledge of ${\cal B}[\td]$ has the advantage of obtaining the characteristic function $F_d(t)$ of the Adler function, where the latter function appears in the integral resummation of the Adler function
\be
d(Q^2)_{D=0; res} = \int_0^{\infty} \frac{dt}{t} F_d (t) a(t Q^2).
\label{dres} \ee
Namely, if the Borel transform ${\cal B}[\td](u)$ fulfills certain convergence conditions, then the characteristic function $F_d(t)$ is the inverse Mellin transform of ${\cal B}[\td](u)$
\be
F_d(t) = \frac{1}{2 \pi i} \int_{u_0-i \infty}^{u_0+i \infty} du \; {\cal B}[\td](u) \; t^u,
\label{Fd1} \ee
where $u_0$ is any real value between the singularities of ${\cal B}[\td](u)$ nearest to the origin, i.e., $-1 < u_0 < +2$. We can take $u_0=+1$, and the change of variable $u \mapsto z$, where $u=1 - i z$, transforms this integral in an integral along the real axis
\be
F_d(t) = \frac{t}{2 \pi} \int_{-\infty}^{+\infty} dz \; {\cal B}[\td](u=1-i z) \; e^{- i z \ln t}.
\label{Fd2} \ee
For the expression (\ref{Btd5P}) for ${\cal B}[\td](u)$, the above formula can be applied without modification to all nonlogarithmic terms, cf.~Appendix \ref{appGd}. For the logarithmic term ($\propto \tal \ln (1 - u/2)$), the convergence conditions are not fulfilled and a subtraction is needed \cite{renmod}. The final resummation formula has the form
\be
d(Q^2)_{D=0; res} = \int_0^{\infty} \frac{dt}{t} G_d (t) a(t e^{-\tK} Q^2) +
\int_0^{1} \frac{dt}{t} G_d^{\tal} (t) \left[ a(t e^{-\tK} Q^2) - a(e^{-\tK} Q^2) \right],
\label{dres5P} \ee
where the characteristic function $G_d^{\tal} (t)$ of the logarithmic part is \cite{renmod}
\be
G_d^{\tal} (t) = - \tal \td^{\rm IR}_{2,1} \frac{\pi \; t^2} {\ln t}.
\label{Gdtal} \ee
The characteristic function $G_d(t)$ for the terms ${\cal B}[\tk](u)_{\xi,\tK} = \pi \exp(\tK u)/(3 -u)^{\xi}$ is given in Appendix \ref{appGd}, Eq.~(\ref{dxitK}). Combining the result (\ref{dxitK}) with the results for the characteristic functions of the simple power terms $1/(2 -u)$ and $1/(1+u)^2$ given in Ref.~\cite{renmod}, we obtain the full expression for the characteristic function $G_d(t)$ appearing in (\ref{dres5P}) and corresponding to the considered Borel transform ${\cal B}[\td](u)$ Eq.~(\ref{Btd5P})
\bes
\label{Gdtot}
\bea
G_d(t) &=& G_d^{(-)}(t) \Theta(1 - t) + G_d^{(+)}(t) \Theta(t - 1),
\label{Gdstr} \\
G_d^{(-)}(t) & = & \pi \left[ \td^{\rm IR}_{2,1} t^2 + \td^{\rm IR}_{3,2} \frac{t^3}{\Gamma(\kappa^{(2)}) \left( \ln(1/t) \right)^{1 - \kappa^{(2)}}}  + \td^{\rm IR}_{3,1} \frac{t^3}{\Gamma(\kappa^{(1)}) \left( \ln(1/t) \right)^{1 - \kappa^{(1)}}} \right],
\label{Gdmi} \\
G_d^{(+)}(t) & = & \pi \td^{\rm UV}_{1,2} \frac{\ln t}{t}.
\label{Gdpl} \eea \ees
However, this resummation has in the considered case of ($\MSbar$) perturbative coupling $a(Q^{'2})$ a serious obstacle, namely the Landau singularities of this coupling at positive $0 < Q^{' 2} \lesssim 1 {\rm GeV}^2$. At sufficiently low positive values of $t$ the integration (\ref{dres5P}) hits these singularities and makes the evaluation there impossible or, at least, ambiguous. Because of this problem, we will not use the resummation (\ref{dres5P}) in the numerical analysis of the sum rules here. The application of such a resummation approach appears to be optimal in the formulations of the QCD where the coupling $a(Q^{'2})$ [$ \mapsto \A(Q^{'2})$] is free of these Landau singularities, i.e., when it is a holomorphic function in the complex $Q^{'2}$-plane except on the negative axis (e.g.~the QCD variant of Refs.~\cite{3dAQCD,amuO}); we intend to pursue this approach in a future work.

\section{Fitting the Borel-Laplace sum rules with ALEPH data}
\label{sec:fitBL}

The analysis was performed with the data for the spectral function Eq.~(\ref{om1}) of the strangeless semihadronic $\tau$-decays of the ALEPH Collaboration \cite{ALEPH2,DDHMZ,ALEPHfin,ALEPHwww} for the (V+A)-channel, cf.~Fig.~\ref{FigOmega}(a), where we used updated values of various relevant parameters and a rescaling factor for the spectral functions as explained in Ref.~\cite{Bo2015}. Furthermore, we also performed analysis of the pure V-channel, which includes data of ALEPH, OPAL and of an additional $\tau$-decay channel and of electroproduction data \cite{Boito:2020xli,Perisetal}, cf.~Fig.~\ref{FigOmega}(b).

For the evaluation of the $D=0$ contribution to the sum rules, we used two methods, each involving a truncation:

\begin{enumerate}
\item Fixed Order Perturbation Theory using powers of $a(\sm)$ ('FO'). This method consists of Taylor-expanding the powers $a(\sm e^{i \phi})^n$, appearing in $d(\sm e^{i \phi})_{D=0}$ in the contour integral of the sum rules, in powers of $a(\sm)$ and\footnote{This is for the (central) cases when using the renormalisation scale parameter value $\kappa=1$. For other values of $\kappa$, the expansion is in powers of $a(\kappa \sm)$; we recall that $\sm (\equiv {\sigma}_{\rm max}) = 2.8 \ {\rm GeV}^2$ in the case of the (V+A)-channel, and $\sm = 3.0574 \ {\rm GeV}^2$ in the case of the V-channel.} truncating the final result at a chosen truncation power index $N_t$, i.e., at $a(\sm)^{N_t}$.
\item Inverse Borel Transformation with Principal Value ('PV'). This method uses the expansion (\ref{Bd5P}) of the Borel transform ${\cal B}[d](u)$ around the renormalon singularities. The expansion is truncated as indicated in Eq.~(\ref{Bd5P}), i.e., the terms indicated there as '${\cal O}(\ldots)$' are not included, and this gives us the ``singular'' part ${\cal B}[d](u)_{\rm sing}$. The expression $d(\sm e^{i \phi})_{D=0}$ is then written as the Principal Value (PV) of the Inverse Borel transform
  \be
{\bigg (} d(\sm e^{i \phi})_{D=0} {\bigg )}^{({\rm PV}, [N_t])} = \frac{1}{\beta_0} \frac{1}{2} \left( \int_{{\cal C}_{+}} + \int_{{\cal C}_{-}} \right) d u \exp \left[ - \frac{u}{\beta_0 a(\sm e^{i \phi})} \right] {\cal B}[d](u)_{\rm sing} + \delta d(\sm e^{i \phi})^{[N_t]}_{D=0}.
\label{PV2}
\ee
The integration paths ${\cal C}_{\pm}$ start at $u=0$ and go above and below the positive real axis in the complex $u$-plane toward ${\rm Re} (u) \to + \infty$ (further details of the paths ${\cal C}_{\pm}$ are irrelevant because of the Cauchy theorem). The expression $\delta d(\sm e^{i \phi})^{[N_t]}_{D=0}$ is a polynomial of the form
\be
\delta d(\sm e^{i \phi})^{[N_t]}_{D=0} = (\delta d)_0 a(\sm e^{i \phi}) + \ldots +(\delta d)_{N_t-1} (a(\sm e^{i \phi}))^{N_t},
\label{deld} \ee
which represents the correction terms needed so that the expansion of the expression (\ref{PV2}), when expanded in powers of $a(\sm e^{i \phi})$, reproduces the terms $d_n (a(\sm e^{i \phi}))^{n+1}$ to the order $(a(\sm e^{i \phi}))^{N_t}$, where $d_n$ are the coefficients of the Adler function predicted by the renormalon-motivated model explained in Sec.~\ref{sec:renmod}. The correction polynomial is needed because the expression for ${\cal B}[d](u)_{\rm sing}$ has its own truncation as explained above. While this correction polynomial $\delta d(Q^2)_{D=0}$ brings in the described evaluation dependence on the truncation index $N_t$, it is expected that this dependence on $N_t$ ($N_t \geq 5$) will be significantly weaker than in the FO approach, because the coefficients $(\delta d)_n$ in the correction polynomials have the largest part of the renormalon growth taken away from them by the inverse Borel transform integral in Eq.~(\ref{PV2}). This will be confirmed in our numerical analysis.
\end{enumerate}

In our previous works \cite{EPJ21,EPJ22} we applied also a third evaluation method, ${\widetilde {\rm FO}}$, which is a variant of the FO method. While in FO the (truncated) expansion of the sum rules is performed in powers of $a(\sm)^n$, in ${\widetilde {\rm FO}}$ it is performed in logarithmic derivatives $\ta_n(\sm)$ defined in Eq.~(\ref{tan}). This method gives results which are considerably less stable under the variation of $N_t$. The main reason for this lies in the fact that $\ta_n(\sm)$ are combinations and products of the beta-function $\beta(a)$ and its derivatives. Since our $\beta(a)$ function is a relatively long perturbative polynomial (5-loop $\MSbar)$, the expressions for $\ta_n(\sm)$ become very long polynomials in $a(\sm)$ when $n$ increases. Furthermore, since $\sm$ is relatively low, $a(\sm)$ is not very small. Therefore, such long perturbative polynomials show numerical instabilities when $n$ (and thus $N_t$) increases. For this reason, we do not present the results of the ${\widetilde {\rm FO}}$ method in this work.

 We now proceed to the fit of the sum rules. The Borel-Laplace sum rules are first fitted with the corresponding ALEPH data for the (V+A)-channel. Subsequently, these sum rules are fit with the V-channel data based on combined ALEPH and OPAL data supplemented by the electroproduction data. In this way, we  obtain, for each method (FO, PV) and each truncation index $N_t$ of the $D=0$ contribution, a set of best-fit values of $\alpha_s$ and of the condensates $\langle O_{2 p} \rangle$. Subsequently, various momenta $a^{(2,n)}(\sm)$ are evaluated, and (for each method) the preferred index $N_t$ is chosen in such a way as to have local stability of the values of these two momenta under the increase by one unit, $(N_t-1) \mapsto N_t$. At such values of $N_t$, we obtain the corresponding central value of $\alpha_s$ and of the condensates for each of the methods. We estimate the theoretical uncertainties of the extracted values by varying $\kappa$, $d_4$, $N_t$ and the number of terms in the OPE sum (\ref{DOPE})-(\ref{DOPEV}). In addition, the experimental uncertainties are also evaluated.

In our fitting of the sum rules (\ref{sr}), we apply the real parts of the (double-pinched) Borel Laplace sum rule Eqs.~(\ref{BL}) and (\ref{Bth})-(\ref{BD6}) 
\be
{\rm Re} B_{\rm exp}(M^2;\sm) = {\rm Re} B_{\rm th}(M^2;\sm),
\label{BLsr} \ee
and for the Borel-Laplace complex scale parameters $M^2$ we take them along three different rays in the first quadrant: $M^2 = |M^2| \exp(i \Psi)$, where $\Psi=0$, $\pi/6$, $\pi/4$, and the lengths of the rays are taken as $0.9 \ {\rm GeV}^2 \leq |M^2| \leq 1.5 \ {\rm GeV}^2$. The choices for these values were discussed in \cite{EPJ21}. In practical evaluations of the fits, we minimised the following sum of squares of the differences of the quantities (\ref{BLsr}):
\be
\chi^2 = \sum_{\alpha=1}^n \left( \frac{ {\rm Re} B_{\rm th}(M^2_{\alpha};\sm) - {\rm Re} B_{\rm exp}(M^2_{\alpha};\sm) }{\delta_B(M^2_{\alpha})} \right)^2 .
\label{chi2} \ee
Here, $M_{\alpha}^2$ was taken as a set of points along the three mentioned chosen rays. Along each of the three rays, we took three equidistant points covering the entire ray. This means that the sum (\ref{chi2}) contained in practice nine terms ($n=9$). Nonetheless, the fit results changed very little when the number of terms was increased. Further, in the sum (\ref{chi2}), the quantities $\delta_B(M^2_{\alpha})$ are the experimental standard deviations of ${\rm Re} B_{\rm exp}(M^2_{\alpha};\sm)$, cf.~Eq.~(\ref{dBM2}) in Appendix \ref{apperr}.

In our fits the weight function $g_{M^2}(Q^2)$, Eq.~(\ref{gM2}), of the Borel-Laplace sum rules contains all powers of $Q^2$ (including the linear term $Q^2$). Consequently, these sum rules are sensitive in principle to the contributions of all the condensates (including $\langle O_4 \rangle$), in addition to the value of $\alpha_s$. This may be regarded as a drawback. However, we believe that this possible drawback is offset by the advantage that these sum rules have an additional (continuous) scale parameter $M^2$ whose variation represents an additional means to improve the reliability of the extracted values of the parameters $\alpha_s$ and of the condensates. On the other hand, the fits using momenta sum rules $a^{(i,j)}(\sm)$ have the advantage that they are sensitive only to a limited number of condensates with anomalous dimension zero (in addition to $\alpha_s$), but they do not have the advantage of having an additional (scale) parameter to vary.

\subsection{(V+A)-channel}
\label{subs:VA}

The experimental data used for the (V+A)-channel are those of ALEPH Collaboration, with the last two bins excluded due to the large experimental uncertainties in those two bins, cf.~Fig.~\ref{FigOmega}(a); we will return to this point later on in this Section. This then gives $\sm (\equiv \sigma_{\rm max})= 2.8 \ {\rm GeV}^2$. 

We first perform minimisation of $\chi^2$ of Eq.~(\ref{chi2}) with the truncated OPE of the Adler function of the form (\ref{DOPE}) with two $D=6$ terms. If we truncate the OPE in such a case at $D=6$, the extracted ranges of values of $\alpha_s$ are reasonable (relatively close to the results obtained and presented later on in this work). Namely, the obtained central values of $\alpha_s(m_{\tau}^2)$ are in such a case approximately $0.318^{+0.008}_{-0.010}$ and $0.320^{+0.006}_{-0.010}$ for the FO and PV methods, respectively, with the optimal $D=0$ sector truncation indices $N_t = 8, 5$, respectively. However, if we include in this analysis $D=8$ OPE term, the values of $\alpha_s(m^2_{\tau})$ increase strongly (by about $0.009$) and the cancellations between the two $D=6$ contributions become very strong. When including OPE terms beyond $D=8$, this cancellation becomes even stronger,\footnote{When not including $D=8$ term, the two $D=6$ contributions have cancellations of about $78 \%$. When including the $D=8$ term, the cancellations of the two $D=6$ terms are about $92 \%$, and the inclusion of $D=10$ term leads to cancellations of about $95 \%$.}
and the extracted value of $\alpha_s$ increases further. This suggests that the method becomes increasingly numerically unstable and gives unrealistic results when including OPE terms beyond $D=6$. Furthermore, the uncertainty $\delta(\alpha_s)_{\rm exp}$ of the extracted $\alpha_s(m_{\tau}^2)$ values due to the experimental uncertainties of the data also increases significantly (by 70 \%) when the term $D=8$ is included, from $\delta(\alpha_s)_{\rm exp}=\pm 0.0023$ to $\pm0.0039$.

For all these reasons, we will take in the OPE Eq.~(\ref{DOPE}) only one term with $D=6$, the formally leading $D=6$ term with $k^{(1)}$ ($\approx 0.222$)
\be
{\cal D}_{\rm th}(Q^2)_{D=6} = \frac{6 \pi^2}{(Q^2)^3} \langle O_6 \rangle_{\rm V+A} \; a(Q^2)^{k^{(1)}}  \qquad (k^{(1)}=0.222),
\label{DD6} \ee
and in the $V$-channel OPE, Eq.~(\ref{DOPEV}), the same expression but with the usual replacement $\langle O_6 \rangle_{\rm V+A} \mapsto 2 \langle O_6 \rangle_{\rm V}$. The use of the simplification Eq.~(\ref{DD6}) in the OPE in principle does not result in a significant change (``error'') with respect to the case of two $D=6$ terms: the new condensate $\langle O_6 \rangle$ is now effectively a combination of the two condensates $\langle O_6^{(1)} \rangle$ and $\langle O_6^{(2)} \rangle$, and the value of $\langle O_6 \rangle$ is small due to the mentioned cancellation effects (we also note that the values $k^{(1)}=0.222$ and  $k^{(2)}=0.625$ of the two power indices are not far from each other). The important advantage of the simplification Eq.~(\ref{DD6}) is the significantly improved numerical stability of the results of the fitting procedure under the extension of the OPE beyond $D=6$.
Our numerical results indicate that in this case, i.e., Eq.~(\ref{DD6}), in the considered (V+A)-channel it is sufficient to truncate the OPE at $D=10$ contribution.

It turns out that the quality of the fit is very good, and that the minimised values of $\chi^2$ Eq.~(\ref{chi2}) are $\sim 10^{-5}$.
The main results of this analysis are presented in Table \ref{tabBL}.
\begin{table}
  \caption{(V+A)-channel: The extracted values of $\alpha_s(m_{\tau}^2)$, and the condensates $\langle O_D \rangle_{\rm V+A}$, in units of $10^{-3} \ {\rm GeV}^D$, as obtained by the described Borel-Laplace sum rule. Included are the optimal truncation numbers ($N_t$) and the values of the fit quality $\chi^2$ [cf.~the text and Eq.~(\ref{chi2})].}
 \label{tabBL}
\begin{ruledtabular}
\begin{tabular}{r|r|rrrr|r|r}
  method & $\alpha_s(m_{\tau}^2)$ &  $\langle O_4 \rangle$ & $\langle O_6 \rangle$ & $\langle O_8 \rangle$ & $\langle O_{10} \rangle$ & $N_t$ & $\chi^2$ \\
\hline
FO                     &  $0.3202^{+0.0122}_{-0.0113}$      & $-4.2^{+4.1}_{-6.0}$ & $+6.0^{+3.4}_{-4.6}$   &  $-2.6^{+3.6}_{-3.3}$  &  $+1.0^{+4.1}_{-4.2}$ & 8 & $2.6 \times 10^{-5}$ \\
PV      &  $0.3216^{+0.0059}_{-0.0107}$      & $-3.0^{+1.4}_{-1.2}$ & $+5.4^{+3.7}_{-2.5}$   &  $-1.8^{+2.3}_{-2.6}$   & $+0.7 \pm 2.8$  &  7 & $1.2 \times 10^{-5}$
\end{tabular}
\end{ruledtabular}

\end{table}
The uncertainties in the values given in Table \ref{tabBL} come from the experimental and various theoretical sources. This is explained in more detail below for the case of the extracted values of the parameter $\alpha_s(m_{\tau}^2)$. The extracted values of $\alpha_s(m_{\tau}^2)$, for each method and with separate uncertainties, are in the considered (V+A)-channel case
\bes
\label{BLresal}
\bea
\alpha_s(m_{\tau}^2)^{\rm (FO)} & = & 0.3202 \pm 0.0028({\rm exp})^{-0.0014}_{+0.0087}(\kappa)^{-0.0087}_{+0.0052}(d_4)^{+0.0028}_{-0.0038}(N_t) \pm 0.0043(O_{14}) \pm 0.0032(N_{\rm bin})
\label{BLalFOa}
\\
& = &  0.3202^{+0.0122}_{-0.0113} \approx 0.320^{+0.012}_{-0.011} \qquad \left( N_t = 8^{+2}_{-3}; \; {\rm V+A} \right), 
\label{BLalFOb}
\\
\alpha_s(m_{\tau}^2)^{\rm (PV)} & = & 0.3216 \pm 0.0027({\rm exp})^{-0.0003}_{+0.0002}(\kappa)^{-0.0095}_{+0.0029}(d_4)^{-0.0001}_{+0.0002}(N_t) \pm 0.0030(O_{14}) \pm 0.0032(N_{\rm bin})
\nonumber\\ &&
\mp 0.0001({\rm amb}) 
\label{BLalPVa}
\\
& = &  0.3216^{+0.0059}_{-0.0107} \approx 0.322^{+0.006}_{-0.011}, \qquad \left( N_t = 7 \pm 2; \; {\rm V+A} \right),
\label{BLalPVb}
\eea
\ees
The above values of $\alpha_s(m^2_{\tau})$ were extracted for the methods FO and PV with the truncation indices $N_t=8,7$, respectively. The choice of these truncation indices $N_t$ is explained later in this Section, by consideration of the local stability of the values of the double-pinched momenta $a^{(2,n)}(\sm)$ ($n=0,1,2$) under the variation of $N_t$.

The uncertainties in Eqs.~(\ref{BLresal}) at the symbol '(exp)' are estimates of the experimental uncertainties originating from the ALEPH data, and were obtained by the method explained in the Appendix of Ref.~\cite{Bo2011}, in which we use the information about the correlation matrix of the (nine) Borel-Laplace transform values ${\rm Re} B_{\rm exp}(M^2_{\alpha};\sm)$. The latter correlation matrix, in turn, uses the correlation matrix of the measured ALEPH values for different bins. In Appendix \ref{apperr} we present the main steps applied in this work to obtain these experimental uncertainties. We refer to Appendix \ref{apperr} for further technical details.
The other uncertainties are of theoretical origin:
\begin{enumerate}
\item
The uncertainties at '($\kappa$)' come from the variation of the renormalisation scale parameter $\kappa \equiv \mu^2/Q^2$ around its central value $\kappa=1$ up to $\kappa=2$ and down to $\kappa=1/2$. The conventions in the separate uncertainties in Eqs.~(\ref{BLresal}) are such that, e.g., in the FO method the value of $\alpha_s(m^2_{\tau})$ decreases by $0.0014$ when $\kappa$ increases from $\kappa=1$ to $\kappa=2$, and increases by $0.0087$ when $\kappa$ decreases from $\kappa=1$ to $\kappa=1/2$.\footnote{Since the scale $|Q^2|=\kappa \sm$ has a very low value of $1.4 \ {\rm GeV}^2$ when $\kappa=1/2$, the evaluation of the coupling $a(Q^2)$ at such $Q^2$ is being influenced by a relative vicinity of Landau singularities (the branching point of such singularities is at $Q^2 \approx 0.43 \ {\rm GeV}^2$), and is thus unreliable.}
\item
The uncertainties at '($d_4$)' come from the variation of the coefficient $d_4$ around its central value $d_4=275$ according to the estimate Eq.~(\ref{d4est}), $d_4 = 275 \pm 63$.
\item
  The uncertainties at '($N_t$)' come from relatively wide variation of the truncation index (in $D=0$ contribution to the sum rule) around its central (i.e., optimal) value; this variation was taken in general up to two units upwards and downwards: in the FO approach, the central value is $N_t=8$, and the interval of variation is $5 \leq N_t \leq 10$.
\item
 The uncertainties at '($O_{14}$)' come when we include in the OPE the additional $D=12$ and $D=14$ contributions. In Table \ref{tabalsVA} we present the extracted values of $\alpha_s(m_{\tau}^2)$, with FO approach, for various cases of truncation of the OPE (and various values of $N_t$). We can see that when $O_{10}$ is included in the OPE, the extracted values start to grow considerably more slowly under the inclusion of an additional term $O_{12}$. However, the growth is not significantly reduced further when we include yet additional terms ($O_{14}$, $O_{16}$). In this context, we mention that the problems of truncation of OPE in sum rules with polynomial (momentum) weight functions, at scales $\sim m_{\tau}^2$, were raised and investigated in Refs.~\cite{Bo2017,Bo2019}.
\item
  The uncertainties '($N_{\rm bin}$)' come when the last two bins of the ALEPH data (i.e., beyond $\sm=2.8 \ {\rm GeV}^2$) are included in the analysis. We decided to keep for the central values the case $\sm = 2.8 \ {\rm GeV}^2$, i.e., without the last two bins, for two reasons: (I) the uncertainties in the last two bins are significantly larger than in other bins [cf.~Fig.~\ref{FigOmega}(a)]; (2) the experimental uncertainties $\delta \alpha_s({\rm exp})$ of the extracted value of $\alpha_s$ increase in this case by more than 50 \%, which indicates that the extracted results become less reliable.
\item
Finally, in the PV approach we have yet another theoretical uncertainty ('amb'), the uncertainty due to the ambiguity of the Borel integration (inverse Borel transform), cf.~\cite{EPJ21} for additional explanation.
\end{enumerate}
\begin{table}
  \caption{(V+A)-channel: The extracted values of $\alpha_s(m_{\tau}^2)$, for FO approach and for different values of truncation of the OPE: $D_{\rm max}$ means that the term with maximal dimension $D=D_{\rm max}$ is included in the OPE. The values are given for various values of the truncation index $N_t$ of the $D=0$ sum rule contribution.}
 \label{tabalsVA}
\begin{ruledtabular}
\begin{tabular}{r|rrrrrr}
  $N_t$ & $D_{\rm max}=6$ & $D_{\rm max}=8$ & $D_{\rm max}=10$ & $D_{\rm max}=12$  & $D_{\rm max}=14$  & $D_{\rm max}=16$ \\
\hline
5 & 0.3103 & 0.3139 & 0.3164 & 0.3176 & 0.3186 & 0.3197\\
6 & 0.3079 & 0.3128 & 0.3167 & 0.3184 & 0.3200 & 0.3214 \\
7 & 0.3074 & 0.3134 & 0.3182 & 0.3204 & 0.3223 & 0.3239 \\
8 & 0.3079 & 0.3148 & 0.3202 & 0.3226 & 0.3245 & 0.3261 \\
9 & 0.3088 & 0.3165 & 0.3221 & 0.3245 & 0.3263 & 0.3278 \\
\end{tabular}
\end{ruledtabular}

\end{table}
Further, the total uncertainties in Eqs.~(\ref{BLresal}) and Table \ref{tabBL} were then obtained by adding all these separate uncertainties in quadrature. It is evident from Eqs.~(\ref{BLresal}) that the theoretical uncertainties are significantly larger than the experimental ones, and that the uncertainty '($d_4$)' [due to the variation of $d_4$, cf.~Eq.~(\ref{d4est})] is usually the dominant one.

\subsection{V-channel}
\label{subs:V}

We repeat the same analysis for the V-channel. We use the combined data of ALEPH, OPAL and of an additional $\tau$-decay channel and of electroproduction data \cite{Boito:2020xli,Perisetal}. The data has 68 bins; in contrast to the (V+A)-channel, the last two bins do not have increased uncertainties. It turns out that a good convergence of the extracted values of $\alpha_s$ is achieved for the first time when terms up to $D=14$ are included. This reflects the known fact \cite{BNP92} that the V-channel in general requires more terms in the OPE. Analogously as in Table \ref{tabalsVA} for the (V+A)-channel case, we present in Table \ref{tabalsV} the extracted values of $\alpha_s(m_{\tau}^2)$ for the V-channel, for various cases of truncation of the OPE (\ref{DOPEV}) (and various values of $N_t$). From the results of Table \ref{tabalsV} we can conclude that, once the $O_{14}$ term is included in the OPE, the extracted values of $\alpha_s(m_{\tau}^2)$ stabilise reasonably under the inclusion of additional OPE terms.\footnote{One may worry that, when the number of terms in the OPE is increased beyond the $O_{14}$-term, the number of parameters to fit becomes larger than seven, while in $\chi^2$ of Eq.~(\ref{chi2}) we used only nine terms ($n=9$). However, if we increase this number to $n=15$ (i.e., five points along each ray in the $M^2$-complex plane), the results of Table \ref{tabalsV} change only little; for $N_t=8$, the extracted values of $\alpha_s(m_{\tau}^2)$ are then (in the case $n=15$): $0.3180$, $0.3129$, $0.3120$, $0.3132$ for $D_{\rm max}=12, 14, 16, 18$.}
  
\begin{figure}[htb] 
\centering\includegraphics[width=90mm]{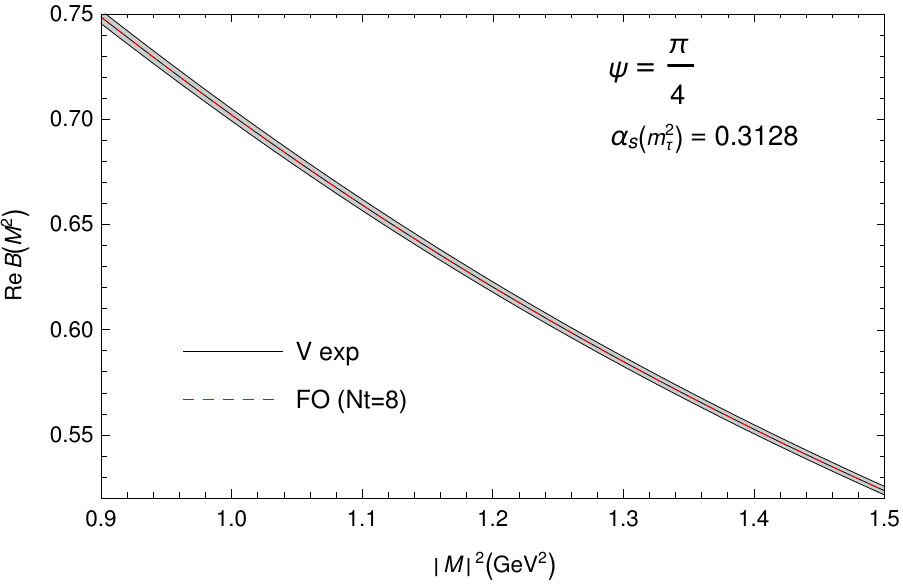}
\caption{\footnotesize (coloured online) The experimental and theoretical values of ${\rm Re} B(M^2; \sm)$ along the ray $M^2=|M^2| \exp(i \pi/4)$, for the V-channel. The (red) dashed line is the result of the FO fit, with $N_t=8$ and the central value $\alpha_s(m_{\tau}^2) =  0.3128$, cf.~Eqs.~(\ref{BLalFOaV})-(\ref{BLalFObV}). The line is virtually indistinguishable from the central experimental line.}
\label{FigPsiPi4}
\end{figure}

\begin{table}
  
  \caption{V-channel: The extracted values of $\alpha_s(m_{\tau}^2)$, for FO approach and for different values of truncation of the OPE. The notations are as in Table \ref{tabalsVA}.}
 \label{tabalsV}
\begin{ruledtabular}
\begin{tabular}{r|rrrrrr}
  $N_t$ & $D_{\rm max}=8$ & $D_{\rm max}=10$ & $D_{\rm max}=12$ & $D_{\rm max}=14$  & $D_{\rm max}=16$ & $D_{\rm max}=18$ \\
\hline
5 & 0.3447 & 0.3236 & 0.3141 & 0.3086 & 0.3079 & 0.3086 \\
6 & 0.3430 & 0.3238 & 0.3148 & 0.3096 & 0.3092 & 0.3101 \\
7 & 0.3437 & 0.3254 & 0.3164 & 0.3113 & 0.3109 & 0.3119 \\
8 & 0.3459 & 0.3274 & 0.3181 & 0.3128 & 0.3124 & 0.3134 \\
9 & 0.3490 & 0.3294 & 0.3196 & 0.3139 & 0.3134 & 0.3143 \\
\end{tabular}
\end{ruledtabular}

\end{table}
The main results of this analysis are presented in Table \ref{tabBLV}.
\begin{table}
    
  \caption{V-channel: The extracted values of $\alpha_s(m_{\tau}^2)$, and the condensates $2 \langle O_D \rangle_{\rm V}$, in units of $10^{-3} \ {\rm GeV}^D$, as obtained by the described Borel-Laplace sum rule. Included is the optimal $N_t$ and the values of the fit quality $\chi^2$ [cf.~the text and Eq.~(\ref{chi2})].}
 \label{tabBLV}
\begin{ruledtabular}
\begin{tabular}{r|r|rrrrrr|r|r}
  method & $\alpha_s(m_{\tau}^2)$ &  $2 \langle O_4 \rangle_{\rm V}$ & $2 \langle O_6 \rangle_{\rm V}$ & $2 \langle O_8 \rangle_{\rm V}$ & $2 \langle O_{10} \rangle_{\rm V}$ & $2 \langle O_{12} \rangle_{\rm V}$ & $2 \langle O_{14} \rangle_{\rm V}$ & $N_t$ & $\chi^2$ \\
\hline
FO      &  $0.3128^{+0.0078}_{-0.0099}$      & $-0.4^{+3.6}_{-3.2}$ & $-19.7^{+2.0}_{-4.9}$   &  $+20.3^{+2.0}_{-2.9}$  &  $-21.4^{+1.7}_{-3.0}$ & $+16.5^{+3.4}_{+2.0}$ & $-7.1^{+9.4}_{-9.5}$ & 8 &   $2.1\times 10^{-11}$ \\
PV      &  $0.3131^{+0.0058}_{-0.0087}$       & $+0.6 \pm 1.0$   &  $-20.8^{+3.9}_{-1.9}$   &  $+21.7^{+3.4}_{-4.0}$  &  $-22.7^{+2.8}_{-2.2}$ & $+17.6^{+3.1}_{+3.5}$ & $-7.6\pm 9.8$ & 7 &   $2.3\times 10^{-11}$
\end{tabular}
\end{ruledtabular}

\end{table}
The extracted values of $\alpha_s(m_{\tau}^2)$, for each method and with separated uncertainties, are in the considered V-channel case
\bes
\label{BLresalV}
\bea
\alpha_s(m_{\tau}^2)^{\rm (FO)} & = & 0.3128 \pm 0.0053({\rm exp})^{-0.0015}_{+0.0038}(\kappa)^{-0.0071}_{+0.0040}(d_4)^{+0.0012}_{-0.0042}(N_t) \mp 0.0004(O_{16})
\label{BLalFOaV}
\\
& = &  0.3128^{+0.0078}_{-0.0099} \approx 0.313^{+0.008}_{-0.010} \qquad \left( N_t = 8^{+2}_{-3}; \; {\rm V-channel} \right), 
\label{BLalFObV}
\\
\alpha_s(m_{\tau}^2)^{\rm (PV)} & = & 0.3131 \pm 0.0053({\rm exp}) \mp 0.0003(\kappa)^{-0.0068}_{+0.0023}(d_4)^{-0.0000}_{+0.0002}(N_t) \mp 0.0007(O_{16}) \mp 0.0001({\rm amb}) 
\label{BLalPVaV}
\\
& = &  0.3131^{+0.0058}_{-0.0087} \approx 0.313^{+0.006}_{-0.009}, \qquad \left( N_t = 7 \pm 2; \; {\rm V-channel} \right),
\label{BLalPVbV}
\eea
\ees
In Fig.~\ref{FigPsiPi4} we present the experimental and theoretical FO values of ${\rm Re} B(M^2; \sm)$ along one of the three rays, namely $M^2=|M^2| \exp(i \pi/4)$, for the V-channel. The narrow grey band represents the experimental values ${\rm Re} B_{\rm exp}(M^2; \sm) \pm \delta_B(M^2)$.
We note that the fit was performed along three different rays simultaneously, with respect to nine points, i.e., three points along each ray (three equidistant points, covering the entire ray between $0.9 \ {\rm GeV}^2 \leq |M^2| \leq 1.5 \ {\rm GeV}^2$).

At the end of Appendix \ref{apperr} we briefly discuss an issue encountered in the evaluation of the experimental uncertainties ('exp') of the extracted parameters in the V-channel case.

\subsection{Determination of optimal truncation index $N_t$}
\label{subs:Nt}

\begin{figure}[htb] 
\begin{minipage}[b]{.49\linewidth}
  \centering\includegraphics[width=85mm]{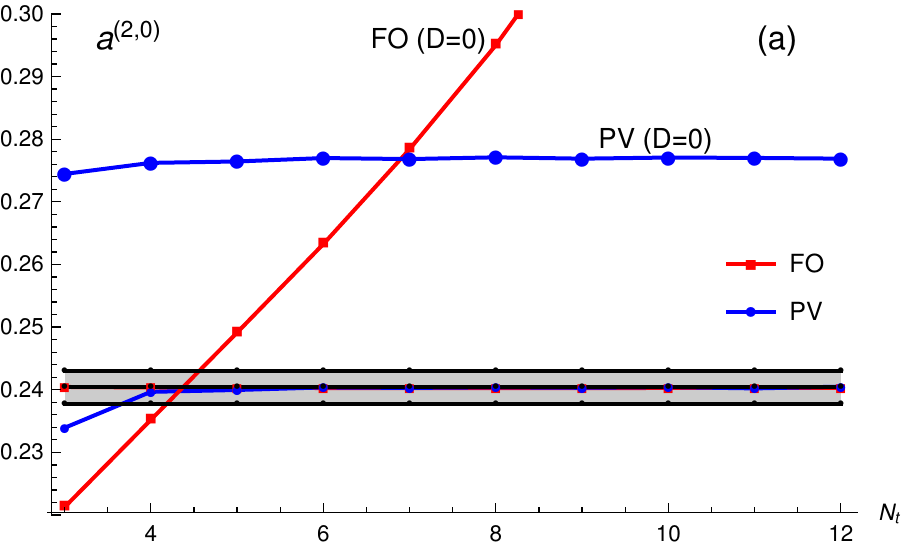}
  \end{minipage} 
\begin{minipage}[b]{.49\linewidth}
  \centering\includegraphics[width=85mm]{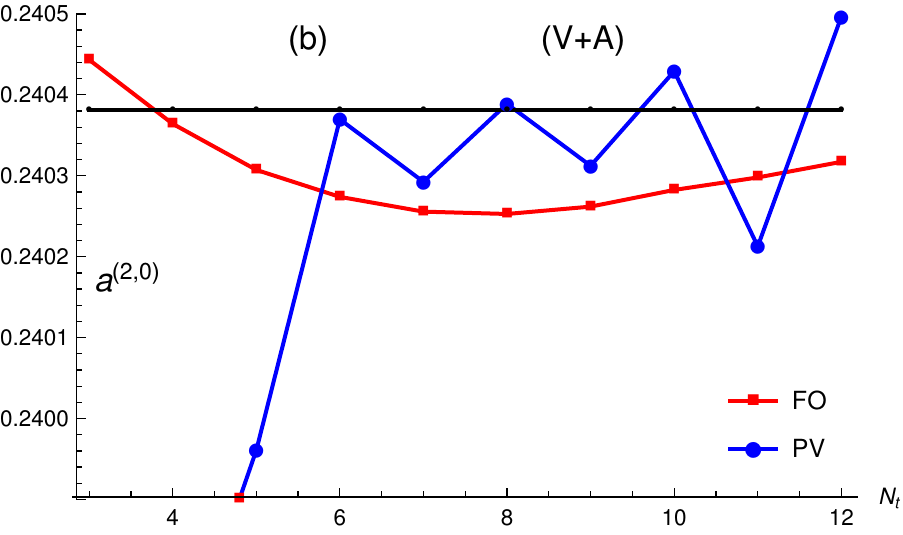}
  \end{minipage} 
\vspace{-0.2cm}
\caption{\footnotesize  (coloured online) (a) The resulting theoretical values of momenta $a^{(2,0)}(\sm)$, for the (V+A)-channel (with $\sm=2.8 \ {\rm GeV}^2$), as a function of the truncation index $N_t$, in the two methods (FO and PV). The grey band is the band of the experimentally allowed values. Included are, for comparison, also the contributions of the $D=0$ part to the momenta. (b) $a^{(2,0)}(\sm)$ on a finer scale of values (where the curves with $D=0$ only are not visible).}
\label{a20avar}
\end{figure}

\begin{figure}[htb] 
\begin{minipage}[b]{.49\linewidth}
  \centering\includegraphics[width=85mm]{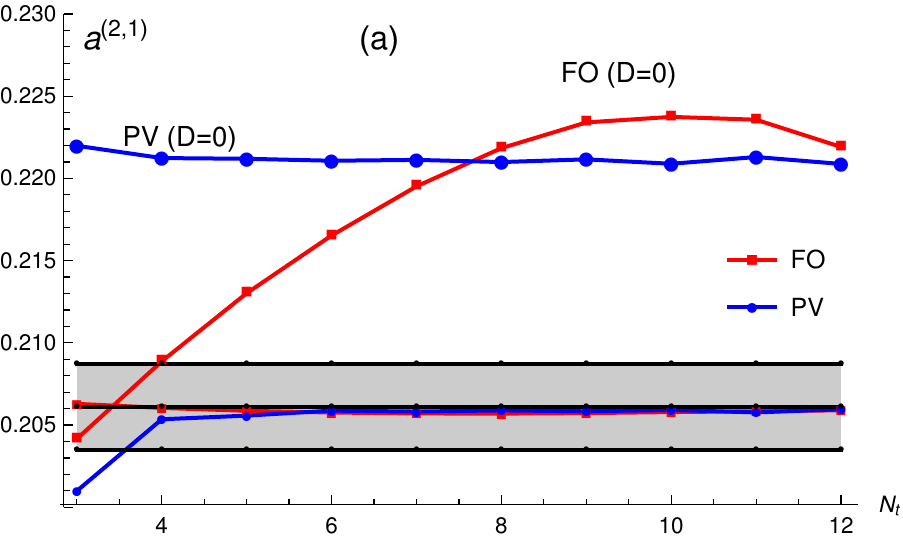}
  \end{minipage} 
\begin{minipage}[b]{.49\linewidth}
  \centering\includegraphics[width=85mm]{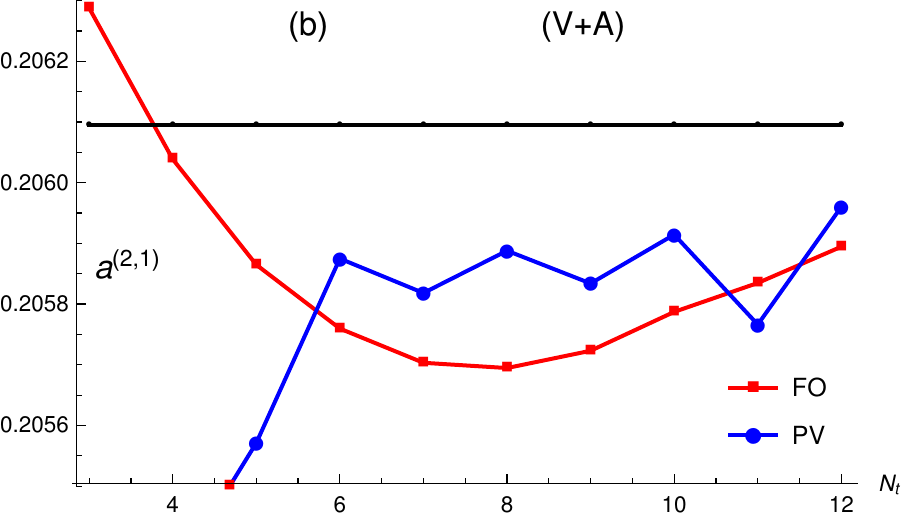}
  \end{minipage} 
\vspace{-0.2cm}
\caption{\footnotesize  (coloured online) The same as Figs.~\ref{a20avar}, but for the momenta  $a^{(2,1)}(\sm)$.}
\label{a21avar}
\end{figure}

\begin{figure}[htb] 
\begin{minipage}[b]{.49\linewidth}
  \centering\includegraphics[width=85mm]{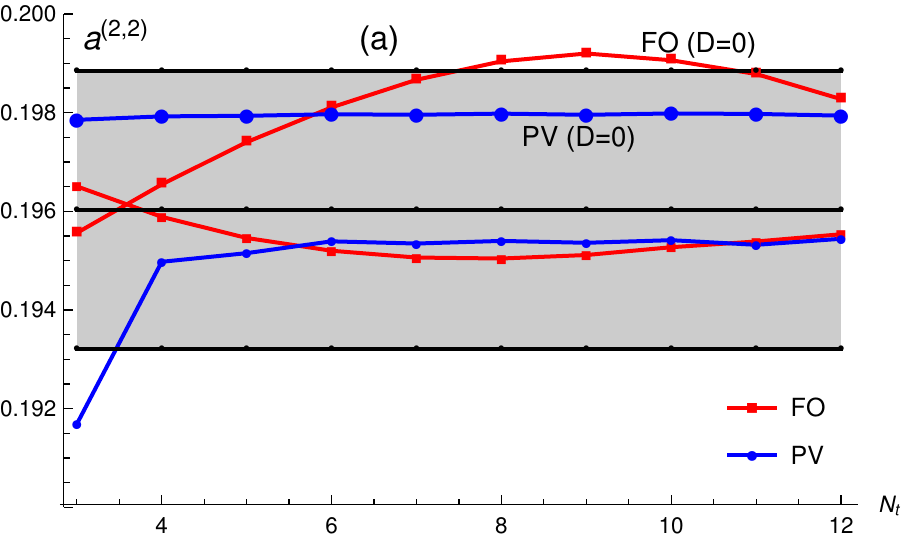}
  \end{minipage} 
\begin{minipage}[b]{.49\linewidth}
  \centering\includegraphics[width=85mm]{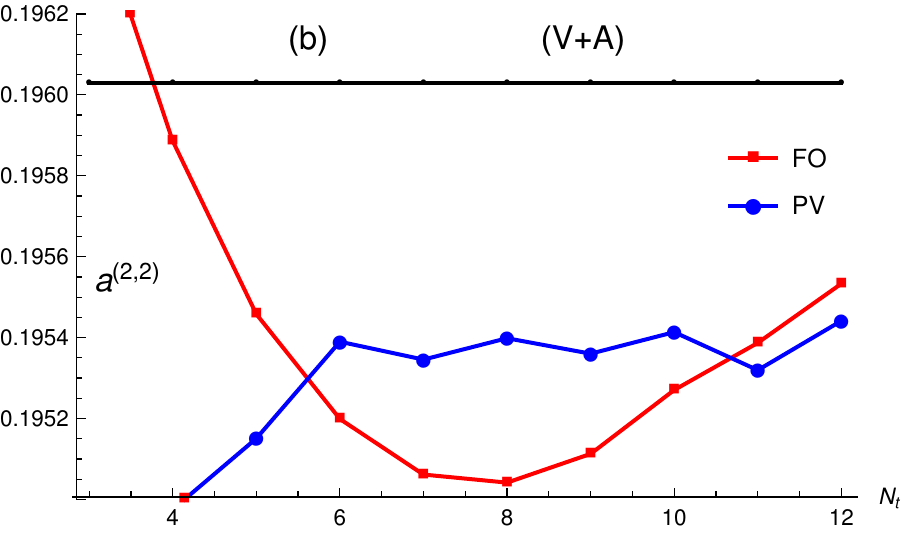}
  \end{minipage} 
\vspace{-0.2cm}
\caption{\footnotesize  (coloured online) The same as Figs.~\ref{a20avar}, but for the momenta  $a^{(2,2)}(\sm)$.}
\label{a22avar}
\end{figure}

As mentioned earlier, the optimal truncation indices $N_t$ of the $D=0$ sum rule contribution for each evaluation method are determined by the momenta $a^{(2,n)}(\sm)$ ($n=0,1,\ldots$) under the variation of $N_t$. We present in Figs.~\ref{a20avar}-\ref{a22avar} the values of the theoretical momenta $a^{(2,n)}(\sm)$ ($n=0,1,2$), Eqs.~(\ref{a2nth}) and (\ref{a2nthexp}), for each truncation index $N_t$, for the two evaluation methods (FO and PV), for the case of the (V+A)-channel (based on ALEPH data).
We point out that Figs.~\ref{a20avar}(b)-\ref{a22avar}(b) present a very narrow range of values of the momenta. This means that both FO and PV methods reproduce, for a wide range of values of $N_t$, to high accuracy the medium experimental values of these momenta. These values are: $a^{(2,0)}(\sm)_{\rm exp}=0.240381\pm 0.002612$;  $a^{(2,1)}(\sm)_{\rm exp}=0.206095\pm 0.002605$; $a^{(2,2)}(\sm)_{\rm exp}=0.196030 \pm 0.002818$.
The results of Figs.~\ref{a20avar}(b)-\ref{a22avar}(b) suggest that, for the FO method, the minimal sensitivity of the theoretically predicted momenta under the increase $(N_t-1) \mapsto N_t$ is reached for the first time at $N_t=8$; and for the PV method at $N_t=7$.

Similar analysis for the V-channel, where the momenta $a^{(2,n)}(\sm)$ ($n=0,1,2,3,4$; $\sm=3.0574 \ {\rm GeV}^2$) are considered, gives the minimal sensitivity of these V-channel momenta at $N_t=8$ for FO and $N_t=7$ for PV (i.e., equal result as for the (V+A)-channel case). In this case, the medium experimental values of these (V-channel) momenta are reproduced to even higher accuracy, because the OPE in our analysis has in this case terms up to the dimension $D=14$.

The results for the momenta $a^{(2,n)}(\sm)$ presented in Figs.~\ref{a20avar}-\ref{a22avar}, are at each truncation index $N_t$, given for the values of $\alpha_s$ and condensates which depend on the index $N_t$ (and on the method: FO, PV). The latter values were obtained by the fit of the Borel-Laplace sum rules to the experimental values at that $N_t$; as a consequence, it can be expected in advance that the resulting values of $a^{(2,n)}(\sm)$ are also within the experimental band and thus relatively stable under the variation of $N_t$. This expectation is confirmed by the results in Figs.~\ref{a20avar}-\ref{a22avar}.

\begin{figure}[htb] 
\begin{minipage}[b]{.49\linewidth}
  \centering\includegraphics[width=85mm]{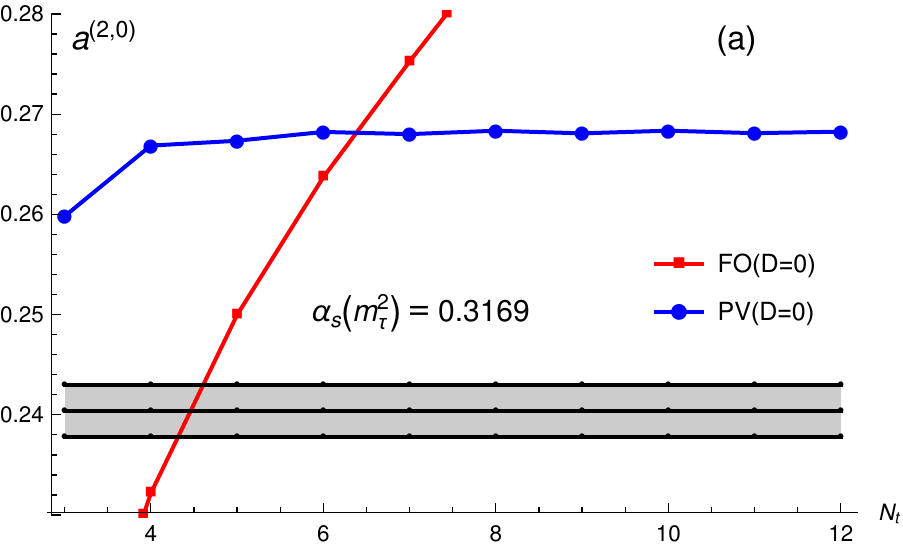}
  \end{minipage} 
\begin{minipage}[b]{.49\linewidth}
  \centering\includegraphics[width=85mm]{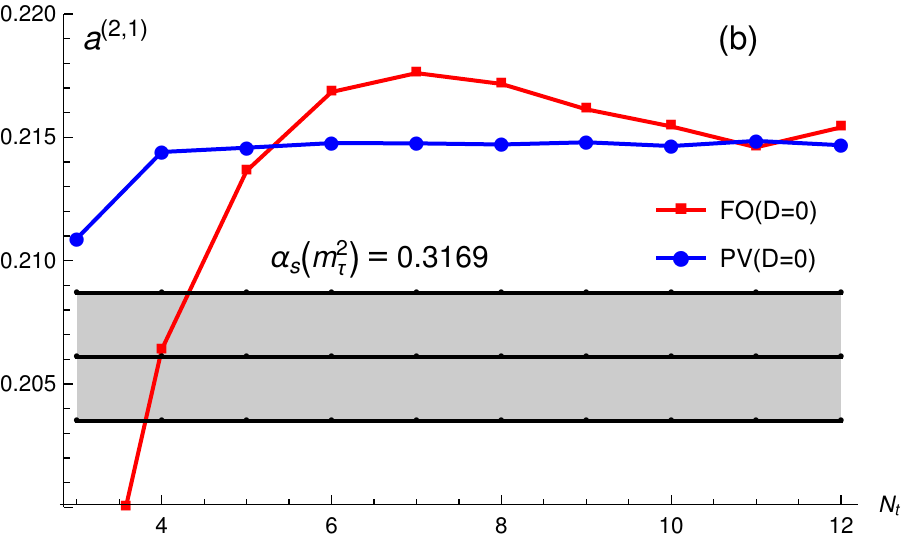}
\end{minipage}
\vspace{-0.2cm}
\caption{\footnotesize  (coloured online) The values of the moment $a^{(2,0)}(\sm)_{(D=0)}$ (a), and $a^{(2,1)}(\sm)_{(D=0)}$ (b), as a function of the truncation index $N_t$, for the two considered methods FO and PV, using for the coupling a fixed value [$\alpha_s(m^2_{\tau})=0.3169$]. The light grey band represents the experimental results for these momenta.}
\label{a2nafixed}
\end{figure}

On the other hand, another question that is interesting in this context is how the expressions for the  momenta $a^{(2,n)}(\sm)$ behave under the variation of $N_t$ when the values of $\alpha_s$ and of the condensates are not varied with $N_t$. In such a case, only the $D=0$ part of the sum rule varies with $N_t$. Therefore, we present the values of the momenta $a^{(2,n)}(\sm)_{D=0}$ ($n=0,1$) in Figs.~\ref{a2nafixed}(a)-(b) for a fixed value of $\alpha_s$, namely the central averaged value of the two methods and the two channels [cf.~Eq.~(\ref{finra})].
As we can see from the comparison of Figs.~\ref{a20avar}-\ref{a22avar}(a)-(b) with those of Figs.~\ref{a2nafixed}, the predictions for the momenta $a^{(2,n)}(\sm)$ in the FO method are much more stable under the variation of $N_t$ when the OPE contributions $D >0$ are included and the values of $\alpha_s$ and of the condensates at each $N_t$ are taken from the corresponding Borel-Laplace sum rule fit. In the PV method (Borel resummation), the stability of the momenta under variation of $N_t$ is moderately good even in the case when $D >0$ terms are not included.

\begin{figure}[htb] 
\begin{minipage}[b]{.49\linewidth}
  \centering\includegraphics[width=85mm]{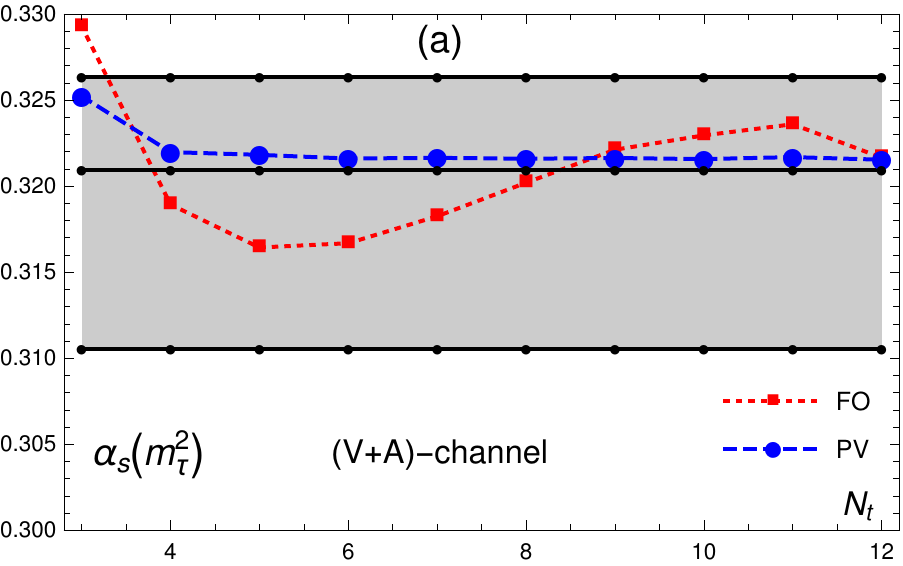}
  \end{minipage} 
\begin{minipage}[b]{.49\linewidth}
  \centering\includegraphics[width=85mm]{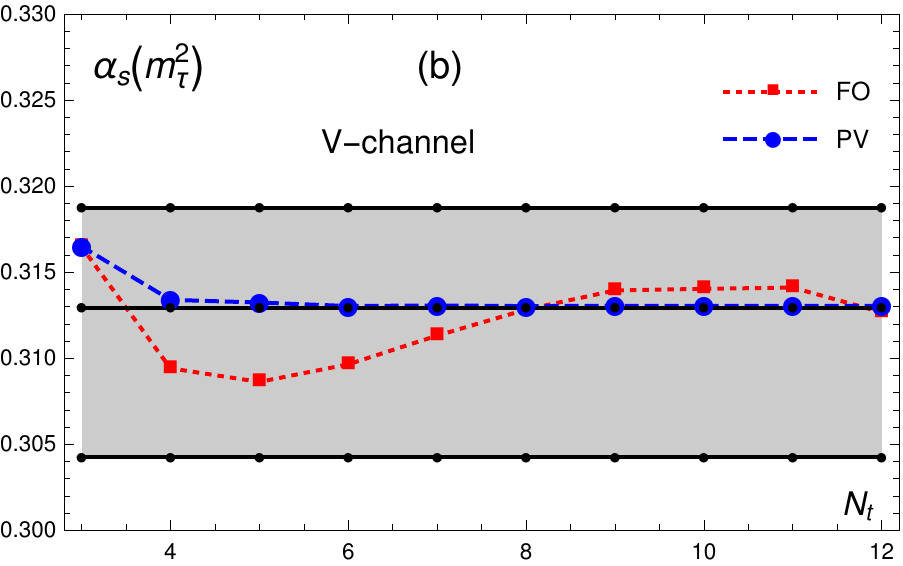}
\end{minipage}
\caption{(coloured online) (a) The values of $\alpha_s(m^2_{\tau})$ extracted from the Borel-Laplace sum rule of the (V+A)-channel, as a function of the truncation index $N_t$, for each of the two methods (FO, PV). The light grey band represents the band of the final averaged values for the (V+A)-channel [cf.~Eq.~(\ref{2avVA})]. (b) The same, but extracted from the V-channel data.}
\label{FigalvsNt}
\end{figure}

\begin{figure}[htb]
  \centering\includegraphics[width=100mm]{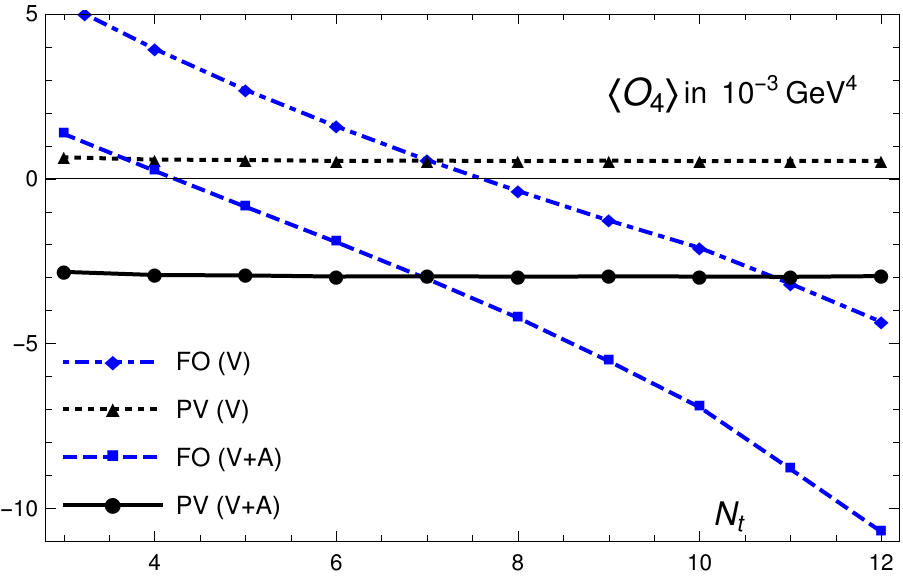}
\vspace{-0.2cm}
\caption{(coloured online) The values of the condensates $\langle O_4 \rangle_{\rm V+A}$ and $2 \langle O_4 \rangle_{\rm V}$, extracted from the Borel-Laplace sum rule from the (V+A) and V-channel data, respectively, as a function of the truncation index $N_t$, for the two methods (FO, PV).}
\label{FigO4vsNt}
\end{figure}

In order to see more clearly how the values of the parameters, extracted from the Borel-Laplace sum rules, depend on $N_t$ in the two methods, in Figs.~\ref{FigalvsNt} we present the extracted values of $\alpha_s(m^2_{\tau})$, and in Fig.~\ref{FigO4vsNt} the extracted values of the (V+A)-channel and V-channel condensates $\langle O_4 \rangle$, as a function of $N_t$, for the two considered methods.

As we can see from all these Figures \ref{a20avar}-\ref{FigO4vsNt}, in practice the PV method Eq.~(\ref{PV2}) appears to be superior to the FO method. An explanation of this is that the major part of the renormalon effects are included in the Borel integral (PV of the inverse Borel transform) in Eq.~(\ref{PV2}), while the $N_t$ dependence is only felt in the truncated correction perturbation series $\delta d(\sm e^{i \phi})^{[N_t]}_{D=0}$ Eq.~(\ref{deld}) that is largely free of the renormalon effects. In Table \ref{tabdvsdeld} we present the values of the coefficients $(\delta d)_n$ of this series vs the coefficients $d_n$ of the (full) $D=0$ Adler function as generated by the renormalon-motivated extension explained in Sec.~\ref{sec:renmod}. In this Table we can see that the coefficients $(\delta d)_n$ become relatively small when $n$ increases, and that the renormalon growth of $d_n$ (with increasing $n$) is less strong in the coefficients $(\delta d)_n$ (this is even more visible when $n > 10$). 
\begin{table}
  \caption{The values of the parameters coefficients $d_n$ and $(\delta d)_n$, for the central case where $d_4=275$ (and $\kappa=1$) and the ratios $(\delta d)_n/d_n$. See the text for further explanation.}
\label{tabdvsdeld}
\begin{ruledtabular}
\begin{tabular}{r|rrrrrrrr}
  $n$: & 3 & 4 & 5 & 6 & 7 & 8 & 9 & 10
\\
\hline
$d_n$: & 49.0757 & 275 & 3206.48 & 16901.6 & 358634. & 621177. & $7.52194 \times 10^{7}$ & $-5.21168 \times 10^{8}$
\\
\hline
$(\delta d)_n$: & 22.0585 & 14.1972 & 255.489 & -661.352 & 11178.3 & -97789.3 & 
$1.1759 \times 10^{6}$ & $-1.34274 \times 10^{7}$
\\
\hline
$(\delta d)_n/d_n$ & 0.44948 & 0.0516262 & 0.0796792 & -0.0391296 & 0.031169 & -0.157426 & 0.0156329 & 0.0257641
\end{tabular}
\end{ruledtabular}
\end{table}

\section{Final results and conclusions}
\label{sec:concl}

In this work we improved on our previous works \cite{EPJ21,EPJ22} of application of the double-pinched Borel-Laplace sum rules to the (V+A)-channel semihadronic strangeless $\tau$ decay data of ALEPH. The $D=0$ Adler function contribution $d(Q^2)_{D=0}$ is based on the Borel transform ${\cal B}[\td](u)$ of an auxiliary Adler function $\td(Q^2)$ according to a renormalon-motivated model \cite{renmod} as in our previous works \cite{EPJ21,EPJ22,Trento}. However, the $u=3$ IR renormalon contribution in ${\cal B}[\td](u)$ is now refined in the sense that it has noninteger power indices, $\sim 1/(3-u)^{\kappa^{(j)}}$ ($j=1,2$) as suggested by the work \cite{BHJ}. Furthermore, the OPE of the full Adler function ${\cal D}(Q^2)$ has now the $D=6$ contribution containing the leading anomalous dimension effect, $\sim a(Q^2)^{k^{(1)}} \langle O_6 \rangle/(Q^2)^3$ with $k^{(1)}=1 - \kappa^{(1)}$ ($\approx 0.222$). All other $D \not=0$ terms in the OPE are taken with zero anomalous dimension. When performing the fit for the (V+A)-channel ALEPH data, we truncate the OPE at $D=10$, to achieve a reasonable stabilisation of the extracted values of $\alpha_s$ under further increase of $D_{\rm max}$. The fit for the V-channel uses combined ALEPH and OPAL data supplemented by electroproduction data \cite{Boito:2020xli,Perisetal}, and we need to truncate the OPE at $D=14$ to achieve the stability of the extracted values of $\alpha_s$. Furthermore, we use for the evaluation of the '(exp)' uncertainties of the extracted values of the parameters ($\alpha_s$ and the condensates), coming from the experimental uncertainties of the data, an approach which accounts correctly for the correlation of the Borel-Laplace sum rules at different scales. This gives us experimental uncertainties $\delta (\alpha_s)_{\rm exp}$ of the extracted values which are significantly higher than those estimated in \cite{EPJ21,EPJ22,Trento}. Nonetheless, our results indicate that in the case of (V+A)-channel (ALEPH data) the theoretical uncertainties of the extracted values of $\alpha_s$, in particular those from the unknown $\sim a^5$ coefficient $d_4$ of the Adler function $d(Q^2)_{D=0}$, are dominant over the experimental uncertainties. On the other hand, in the case of the V-channel data (combined ALEPH and OPAL data), the experimental and theoretical uncertainties are mutually comparable.

The truncation $N_t$ index in the $D=0$ part was fixed in a way similar as in our previous works \cite{EPJ21,EPJ22}, namely by considering the resulting values of the (double-pinched) momenta sum rules $a^{(2,n)}(\sm)$ ($n \leq 2$ for V+A; $n \leq 4$ for V) at each $N_t$ for each method, and choosing the value of $N_t$ where these momenta become locally almost independent of the variation of $(N_t-1) \mapsto N_t$. Particularly the PV method resulted in very little dependence on the variation of $N_t$ (for $N_t \geq 4$).

The duality violations (DV) were suppressed in our approach in two ways: all the weight functions $g(Q^2)$ in the applied sum rules (\ref{sr}) are double-pinched,\footnote{The corresponding weight functions $G(Q^2)$ Eq.~(\ref{GQ2}) are then triple-pinched.} i.e., they have double zero in the timelike point $Q^2 = -\sigma_{\rm max}$ ($\equiv -\sm$). Further, in the Borel-Laplace weight functions there is an additional exponential suppression factor $\exp(-\sm/M^2)$ in this point $Q^2 = -\sm$.

The main results, were presented in Eqs.~(\ref{BLresal}) and Table \ref{tabBL} for the (V+A)-channel, and in Eqs.~(\ref{BLresalV}) and Table \ref{tabBLV} for the V-channel. The arithmetic average of the results over the two methods, for each channel case, gives
\bes
\label{2av}
\bea
\alpha_s(m_{\tau}^2)_{\rm {V+A}} &=& 0.3209^{+0.0059}_{-0.0107}, 
\label{2avVA} \\
\alpha_s(m_{\tau}^2)_{\rm {V}} &=& 0.3129^{+0.0058}_{-0.0087}.
\label{2avV} \eea \ees
The uncertainties $^{+0.0059}_{-0.0107}$ in Eq.~(\ref{2avVA}) were obtained by adding in quadrature the deviation between the average value $0.3209$ and the central value of one of the two methods ($\pm 0.0007$) [cf.~Eqs.~(\ref{BLalFOb}) and (\ref{BLalPVb})], and the uncertainties of the method which gives the smallest uncertainties among the two methods ($^{+0.0059}_{-0.0107}$) [cf.~Eq.~(\ref{BLalPVb})], similar to the reasoning in Refs.~\cite{Pich,EPJ21}. The uncertainties $^{+0.0058}_{-0.0087}$ in Eq.~(\ref{2avV}) were obtained analogously, using Eqs.~(\ref{BLresalV}). It is interesting that the deviation between the average value and the central of (any) one of the two methods is so small that it does not affect the uncertainties in Eqs.~(\ref{2av}), neither in the case of the (V+A)-channel nor in the case of the V-channel.

We perform the average of the two results Eqs.~(\ref{2av}) in analogous way, leading to our final result
\bes
\label{finr}
\bea
\alpha_s(m_{\tau}^2) &=& 0.3169^{+0.0070}_{-0.0096},
\label{finra} \\
\Rightarrow \;
\alpha_s(M_{Z}^2) &=& 0.1183^{+0.0009}_{-0.0012}.
\label{finrb} \eea \ees
Observing Eqs.~(\ref{BLresal}), we see that in the (V+A)-channel the total uncertainties are dominated by theoretical uncertainties. On the other hand, observing Eqs.~(\ref{BLresalV}), we see that in the V-channel the experimental uncertainties are larger and are comparable with the theoretical uncertainties. In this context, it would be very helpful to know the exact value of the coefficient $d_4$ (at $a^5$) of the Adler function $d(Q^2)_{D=0}$, as the variation (\ref{d4est}) in the estimate of $d_4$ is the largest source of the theoretical uncertainties in this analysis.

We present, for comparison, in Table \ref{tabrescomp}, the values of $\alpha_s(m_{\tau}^2)$ extracted from the ALEPH $\tau$-decay data by various authors or groups of authors who used specific sum rules and/or methods of evaluation.\footnote{\label{ast} $^{*}$ The results in our previous work \cite{EPJ21} were obtained for the $d_4$-parameter range $d_4=338 \pm 63$, but in Table \ref{tabrescomp} they were adjusted to the range $d_4=275 \pm 63$ used in the later works \cite{EPJ22,Trento} and here, in order to make comparisons with those works and the present work clearer.}
\begin{table}
  \caption{The values of $\alpha_s(m_{\tau}^2)$ as extracted by applying sum rules to the ALEPH $\tau$-decay data and various methods of evaluation. In the second column, 'BL' stands for Borel-Laplace sum rules, 'DV' for a Duality Violation model in sum rules. For the entry \cite{EPJ21}$^{*}$ below, see footnote \ref{ast}. The last entry contains our results for the $V$-channel which contains combined ALEPH, OPAL and electroproduction data.}
 \label{tabrescomp}
\begin{ruledtabular}
\begin{tabular}{l|l|lll|l}
group &  sum rule & FO & CI & PV & average \\
\hline
Baikov et al.~\cite{BCK} & $a^{(2,1)}=r_{\tau}$ & $0.322 \pm 0.020$ & $0.342 \pm 0.011$ & --- & $0.332 \pm 0.016$ \\
Beneke \& Jamin \cite{BJ} & $a^{(2,1)}=r_{\tau}$ & $0.320^{+0.012}_{-0.007}$ & --- & $0.316 \pm 0.006$ & $0.318 \pm 0.006$ \\
Caprini \cite{Caprini2020} & $a^{(2,1)}=r_{\tau}$ &  --- & --- & $0.314 \pm 0.006$ &  $0.314 \pm 0.006$ \\
Davier et al.~\cite{ALEPHfin} & $a^{(i,j)}$ & $0.324$ & $0.341 \pm 0.008$ & --- & $0.332 \pm 0.012$ \\
Pich \& R.-S.~\cite{Pich}   &  $a^{(i,j)}$     & $0.320 \pm 0.012$ &  $0.335 \pm 0.013$ & --- & $0.328 \pm 0.013$  \\
Boito et al.~\cite{Bo2015} & DV in $a^{(i,j)}$ & $0.296 \pm 0.010$ & $0.310 \pm 0.014$ & --- & $0.303 \pm 0.012$ \\
our previous work \cite{EPJ21}$^{*}$  & BL & $0.311 \pm 0.007$ & $0.341^{+0.010}_{-0.007}$ & $0.320^{+0.009}_{-0.006}$ & $0.315 \pm 0.009$ (FO+PV; V+A) \\
our previous work \cite{EPJ22,Trento}   & BL & $0.323^{+0.013}_{-0.012}$ & $0.349^{+0.014}_{-0.003}$ & $0.327^{+0.027}_{-0.009}$ & $0.324 \pm 0.013$
(FO+PV; V+A) \\
this work & BL & $0.320^{+0.012}_{-0.011}$ & ---  & $0.322^{+0.005}_{-0.010}$ & $0.321^{+0.005}_{-0.010}$
(FO+PV; V+A) \\
this work & BL & $0.313^{+0.008}_{-0.0010}$ & --- & $0.313^{+0.006}_{-0.009}$ & $0.313^{+0.006}_{-0.009}$
(FO+PV; V)
\end{tabular}
\end{ruledtabular}
\end{table}
For a discussion of most of the results in this Table, we refer to \cite{EPJ21}.

Since our results have a relatively high truncation index $N_t=7$ or $8$ (where $a^{N_t}$ is the power in the $D=0$ contribution where the truncation is made), one may wonder how much the obtained results depend on the specific renormalon-motivated Adler function. The FO approach is independent of the renormalon-motivated extension for $N_t=5$ [the coefficient $d_4$ is taken according to Eq.~(\ref{d4est})]. This is not quite true for the PV approach, where the singular part of the Borel transform ${\cal B}[d](u)$ is resummed to all powers. Therefore, at least for the FO approach, the variation of the extracted value of $\alpha_s$ when $N_t$ varies from the (optimal) $N_t=8$ down to $N_t=5$ ($=8-3$) can be regarded as a measure of the model-dependence of our results. We can see from the uncertainties '($N_t$)' in the results Eqs.~(\ref{BLalFOa}) and (\ref{BLalFOaV})\footnote{The lower variation there, i.e., $-0.0038$ and $-0.0042$ in the (V+A) and V cases, respectively.} that this dependence, although sizable, is not the dominant one.

The numerical results were obtained by programs written in Mathematica, they are freely available \cite{prgs} and include the data for the sum rules based on the ALEPH (V+A)-channel data. The data for the V-channel (the combined ALEPH and OPAL data supplemented by the electroproduction data) can be obtained from \cite{Perisetal}.

We believe that some elements of the methods applied here can be applied also to the Light-Cone Sum Rules (LCSRs). LCSRs involve partial resummations in the three-point QCD sum rules and can be used to evaluate form factors and other hadronic quantities (cf.~Refs.~\cite{BaBrKo,BrFi,ChZh,BeKh,Kh,BMPS} for some earlier works on the subject). If we have some knowledge of the renormalon structure of the spacelike quantities (correlators) appearing in a specific LCSR, then the PV method of resummation described in the present work could possibly be applied.   

\begin{acknowledgments}
This work was supported in part by FONDECYT (Chile) Grants No.~1200189 and No.~1220095.
\end{acknowledgments}

\appendix

\section{Borel transforms of the UV $u=-1$ renormalon contribution}
\label{appUV1}

The expressions for the UV renormalon singularities at $u=-p$ ($<0$), in the Borel transforms ${\cal B}[\td]$ and ${\cal B}[d]$, are analogous to those of the IR renormalon singularities at $u=p$ ($>0$) Eqs.~(\ref{BtdpkvsBdpk})
\bes
\label{BtdpkvsBdpkUV}
\bea
{\cal B}[\td](u)_{-p, \kappa_{-p}^{(j)}} &=& \frac{\pi \td^{\rm UV}_{p,j}}{(p+ u)^{\kappa_{-p}^{(j)}}}
\; \Rightarrow
\label{BtdpkUV}
\\
{\cal B}[d](u)_{-p, \kappa_{-p}^{(j)}} &=& \frac{\pi d^{\rm UV}_{p,j}}{(p+ u)^{\kappa_{-p}^{(j)}- p c_1/\beta_0}} {\Big \{} 1 + \frac{(b_1^{(- 2 p)} + {\cal C}^{(-2 p)}_{1,j})}{(-\beta_0) (\kappa_{-p}^{(j)}- p c_1/\beta_0 -1)} (p + u) +
\nonumber\\
&& + \frac{(b_2^{(-2 p)} + b_1^{(-2 p)} {\cal C}^{(-2 p)}_{1,j} + {\cal C}^{(-2 p)}_{2,j})}{(-\beta_0)^2  (\kappa_{-p}^{(j)}- p c_1/\beta_0 -1) (\kappa_{-p}^{(j)}- p c_1/\beta_0 -2)} (p + u)^2 + \cdots {\Big \}}.
\label{BdpkUV} \eea \ees
The expressions $b_n^{(-2 p)}$ are given in Eqs.~(\ref{b1b2}), by replacing there $p \mapsto -p$.
The leading UV renormalon has $-p=-1$. In the large-$\beta_0$ (LB) approximation the power indices are $\kappa_{-1}^{(2)}=2$ and $\kappa_{-1}^{(1)}=1$ \cite{LB1,LB2,ren}. We include in our approach in the Borel transform ${\cal B}[\td](u)$ Eq.~(\ref{Btd5P}) the leading UV renormalon $-p=-1$ with only one term [cf.~Eq.~(\ref{BtdpkUV})] and use there the power index $\kappa_{-1}=2$, i.e., the leading term in the LB approximation.

\section{Results for the subleading cases of renormalon contributions}
\label{appSL}

As can be seen from the transformation (\ref{tdnkap}), the change of the renormalisation scale parameter $\kappa$ generates from the one term expression of the type Eq.~(\ref{Btdpk}) (at $\kappa=1$) a series of terms with power indices $\kappa_p^{(j)}$, $\kappa_p^{(j)}-1$, $\kappa_p^{(j)}-2$, etc., in ${\cal B}[\td](u;\kappa)$. This then implies the generation of the corresponding terms in ${\cal B}[d](u; \kappa)$ of the type (\ref{Bdpk}), which includes the cases of these lower indices $\kappa_p^{(j)}-1$, $\kappa_p^{(j)}-2$, etc. In Table \ref{tabratCSL} we display the numerical results for these subleading cases, in analogy with the results of Table \ref{tabratC} where only the leading cases were displayed.
\begin{table}
  \caption{As in Table \ref{tabratC}, but now including the cases of subleading (SL) indices $\kappa_p^{(j)}-1$ (SL), $\kappa_p^{(j)}-2$ (SSL), etc. The SL case for UV $(-p=-1, j=2)$ was already given in Table II of \cite{renmod}, and the subleading (SL) case of IR $p=2, j=0$ in Table 1 of \cite{EPJ21}. The notations SL and SSL mean that the power index $\kappa_{\pm p}^{(j)}-n$ in ${\cal B}[\td](u)$ is  $\kappa_{\pm p}^{(j)}-1$ and $\kappa_{\pm p}^{(j)}-2$, respectively.}
\label{tabratCSL}
\begin{ruledtabular}
\begin{tabular}{llll|r|rr}
  type (X) & $\pm p$ & $j$ & $\kappa_{\pm p}^{(j)}-n$ & $d_{p,k}^{\rm X}/\td_{p,k}^{\rm X}$ &  ${\cal C}^{(D)}_{1,j}$ & ${\cal C}^{(D)}_{2,j}$
\\
\hline
X=IR & $p=2$ & $j=0$ SL & $(0-1)$ & $(+2.00 \pm 0.02)$ & $(+17. \pm 1.)$ & $\cdots$
\\
\hline
X=IR & $p=3$ & $j=1$ SL & $(0.778-1)$ &  $(-0.0455 \pm 0.0017)$ & $(+6.96 \pm 2.24)$ & $(+56.8 \pm 13.3)$
\\
X=IR & $p=3$ & $j=1$ SSL & $(0.778-2)$ &  $(+0.727 \pm 0.064)$ & $(+14.58 \pm 5.62)$ & $(+118.2 \pm 30.4)$
\\
X=IR & $p=3$ & $j=2$ SL & $0.375-1$ & $(-0.807 \pm 0.042)$ & $(+9.99 \pm 3.17)$ & $(+76.0 \pm 18.2)$
\\
\hline
X=UV & $-p=-1$ & $j=2$ SL & $2-1$ & $(+5.0098 \pm 0.0001)$ & $(+0.0 \pm 0.0)$ & $(+0.5 \pm 0.1)$
\end{tabular}
\end{ruledtabular}
\end{table}
The UV SL case ($-p=-1; j=2$ SL)  was already given in Table II of \cite{renmod}, and the IR SL ($p=2, j=0$ SL) in Table 1 of \cite{EPJ21}.\footnote{There is a minor typo there, the ratio $d^{\rm IR}_{2,-1}/\td^{\rm IR}_{2,-1}$ there is given as $(+2.00 \pm 0.002)$, it should be $(+2.00 \pm 0.02)$.}
We recall that the IR case $(p=2, j=0)$ is given in Table \ref{tabratC} and corresponds to ${\cal B}[\td](u) = \pi \td^{\rm IR}_{2,0} (-1) \ln(1 - u/2)$; while the SL case of this, given here, corresponds to ${\cal B}[\td](u) = \pi \td^{\rm IR}_{2,-1} (2-u) \ln(1 - u/2)$. The IR case $(p=2, j=1)$ is given in Table \ref{tabratC} and corresponds to ${\cal B}[\td](u) = \pi \td^{\rm IR}_{2,1}/(2-u)^1$; the SL case of this is a simple constant for ${\cal B}[\td](u)$ and thus for ${\cal B}[d](u)$.  Similary, the UV case of $(-p=-1, j=2)$ has ${\cal B}[\td](u) = \pi \td^{\rm UV}_{-1,2}/(1+u)^2$, and the SSL case of this is a simple constant for ${\cal B}[\td](u)$ and thus for ${\cal B}[d](u)$.

\section{Characteristic function $G_d$}
\label{appGd}

In Ref.~\cite{renmod}, the characteristic function $G_d(t)$ appearing in Eq.~(\ref{dres5P}) was obtained for the terms of ${\cal B}[\td](u)$ of the IR renormalon form $\exp(\tK u)/(p-u)^k$ for integer powers $k=2,1$; as well as for the UV renormalon form $\exp(\tK u)/(1+u)^2$.

Apart from these terms, in the presently considered form Eq.~(\ref{Btd5P}) for ${\cal B}[\td](u)$ we also have terms of the form $\exp(\tK u)/(3-u)^{\xi}$ where: $\xi= \kappa^{(1)} = 0.778$; or $\xi = \kappa^{(2)} = 0.375$.
For simplicity, we first omit the exponential factor $\exp(\tK u)$ in these terms ($\tK \mapsto 0$); this factor will be incorporated in a trivial way later. This means that our starting Borel transform is
\be
{\cal B}[\td](u)_{\xi}  \equiv \frac{\pi}{(3-u)^{\xi}}.
\label{Btdxi} \ee
We use the form Eq.~(\ref{Fd2}) of the inverse Mellin transform (i.e., with: $u= 1 - i z$), to obtain the corresponding characteristic function $G_d(t)_{\xi}$
\be
G_d(t)_{\xi} = \frac{t}{2 \pi} \int_{-\infty}^{+\infty} dz \; {\cal B}[\td](1 - i z)_{\xi}  \; e^{-i z \ln t} = \frac{t}{2} {\cal J}_{\xi}(t),
 \label{Gdxi1} \ee
 where
 \be
{\cal J}_{\xi}(t) = \int_{-\infty}^{+\infty} \frac{dz}{(2 + i z)^{\xi}} \; e^{- i z \ln t}.
\label{Jxi1} \ee
The cut of the function ${\cal B}[\td](u)_{\xi}$, Eq.~(\ref{Btdxi}), is along the real semiaxis $3 \leq u$. This then corresponds to the cut $2 \leq (- i) z$ of the function ${\cal J}_{\xi}(t)$, Eq.~(\ref{Jxi1}), i.e., along the positive imaginary semiaxis in the $z$-plane.

\begin{figure}[htb] 
  \centering\includegraphics[width=60mm]{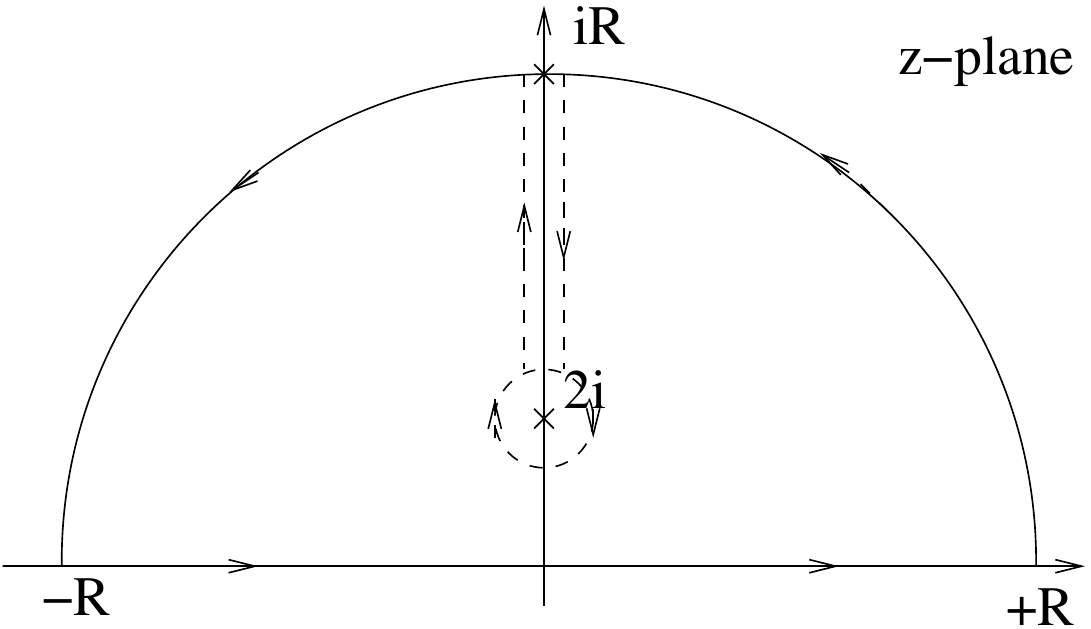}
\vspace{-0.2cm}
\caption{\footnotesize  The closed contour of integration in the complex $z$-plane for the integrand $(e^{- i z \ln t})/(2 + i z)^{\xi}$ for $0<t<1$ ($\ln t < 0$). We have $R \to + \infty$. Around the point $z= 2 i$ there is a circular path of radius $\delta$. The vertical lines of integration have $z= i (2 + s) \pm \epsilon$ where $s$ runs across the interval $\delta < s < (R-2)$. The limiting process is $0 < \epsilon \ll \delta \to +0$.}
\label{Jxicont}
\end{figure}

When $t >1$, we have $\ln t > 0$ and we can close the contour in the lower semicircle, giving us zero value of ${\cal J}_{\xi}(t)$ in such a case. However, when $0 < t < 1$ ($\ln t < 0$), we apply Cauchy theorem to the contour of integration presented in Fig.~\ref{Jxicont}, the contour which does not encircle the singularities of the integrand.
Performing this integration carefully, and applying the limiting procedures mentioned in the caption of the Figure, we obtain (for $0 < t < 1$)
\bes
\label{Jxi2all}
\bea
 {\cal J}_{\xi}(t) &=& 2 t^2 \sin(\xi \pi) \int_0^{+\infty} ds s^{-\xi} t^s
 \label{Jxi2a} \\
 &=& 2 t^2 \sin(\xi \pi) \frac{\Gamma(1 - \xi)}{(\ln (1/t))^{1 - \xi}}
 \label{Jxi2b} \\
 &=& 2 t^2 \frac{\pi}{\Gamma(\xi)} \frac{1}{(\ln (1/t))^{1 - \xi}} \quad (t<1).
 \label{Jxi2} \eea \ees
 The integral over $s$ in Eq.~(\ref{Jxi2a}) converges only for $\xi < 1$. However, the resulting expression (\ref{Jxi2}) exists also for $1 \leq \xi$, it represents an analytic continuation in $\xi$ and is valid also for $1 \leq \xi$.
 This, together with the relation (\ref{Gdxi1}), then gives us the result
\be
G_d(t)_{\xi} =  \frac{\pi}{\Gamma(\xi)} \frac{t^3}{(\ln (1/t))^{1 - \xi}} \Theta(1 - t).
\label{Gdxi2} \ee
Here, the Heaviside step function $ \Theta(1 - t)$ is unity for $t <1$ and zero for $t>1$.

This result can be extended to the case of the Borel transform recaled by an exponential factor
\be
{\cal B}[\td](u)_{\xi,\tK}  \equiv \exp(\tK u) \frac{\pi}{(3-u)^{\xi}}.
\label{BtdxitK} \ee
Namely, by defining $t' \equiv t \exp(\tK)$, we obtain for the characteristic function the same result, except that now $t$ is replaced by $t'$. We can then write the resummation integral [the first integral on the right-hand side of Eq.~(\ref{dres5P})] in terms of $t'$ and renaming back $t' \mapsto t$, and we thus obtain for the Borel transform ${\cal B}[\td](u)$ of the form (\ref{BtdxitK}) the corresponding resummed ``Adler'' function
\bes
\label{dxitK}
\bea
d(Q^2)_{\xi, \tK}  &=& \int_0^{+\infty} \frac{dt}{t} G_d(t)_{\xi} a(t e^{-\tK} Q^2)
\label{dxitKa} \\
& = & 
\frac{\pi}{\Gamma(\xi)} \int_0^{+1} \frac{dt}{t} \frac{t^3}{(\ln (1/t))^{1 - \xi}} a(t e^{-\tK} Q^2).
\label{dxitKb} 
\eea \ees
The same calculation can be repeated for the case when the Borel transform in ${\cal B}[\td](u)_{\xi}$ in Eq.~(\ref{Btdxi}) has $1/(2 - u)^{\xi}$ [instead of: $1/(3 - u)^{\xi}$]; the result for $G_d(t)$ is the same as in Eq.~(\ref{Gdxi2}), except that the factor $t^3$ gets replaced by $t^2$.

\section{Evaluation of experimental uncertainties of extracted parameters}
\label{apperr}

First we explain how the covariance matrix of the Borel sum rules ${\rm Re} B(M^2)$ is obtained. The covariance matrix elements $U_{ij} \equiv U(\sigma_i,\sigma_j)$ for the experimental $\omega(\sigma)$ spectral functions (i.e., here either $\omega_{\rm V+A}$, or $\omega_{\rm V}$) are the following expectation values:
\bea
U_{ij} \equiv U(\sigma_i,\sigma_j) & \equiv & \langle \Delta \omega(\sigma_i) \Delta \omega(\sigma_j) \rangle
\label{covM}
\eea
where 
\be
\Delta \omega(\sigma_i)  \equiv \omega(\sigma_i) - \langle \omega(\sigma_i) \rangle.
\label{DOm}
\ee
Here $\langle \omega(\sigma_i) \rangle = \omega_{\rm exp}(\sigma_i)$ is the experimental average (central) value for the measurements of $\omega(\sigma_i)$ in the i'th $\sigma$-bin, i.e., the bin whose length is $(\Delta \sigma)_i$ and its central point is $\sigma_i$. These matrices for the V-channel are available in the corresponding data \cite{Perisetal}, and for the (V+A)-channel are extractable from the ALEPH data \cite{ALEPHfin}.

The covariance matrix $U_B$ for the Borel sum rules ${\rm Re} B(M^2)$ is then obtained from the evaluation of experimental value of ${\rm Re} B(M^2)$ as a sum over bins [cf.~left-hand side of Eq.~(\ref{sr})]
\be
{\rm Re} B(M^2) = \sum_{j=1}^N (\Delta \sigma)_j f(\sigma_j;M^2) \omega(\sigma_j)
\label{ReBexp}
\ee
where [cf.~Eq.~(\ref{gM2})]
\be
f(\sigma_j;M^2) \equiv {\rm Re} \; g_{M^2}(-\sigma_j)
= {\rm Re} \left( 1 - \frac{\sigma_j}{\sm} \right)^2 \frac{1}{M^2} \exp \left( - \frac{\sigma_j}{M^2} \right)  .
\label{fsj}
\ee
The Borel sum rule covariance matrix $(U_B(\Psi))_{\alpha \beta} \equiv U_B(M^2_{\alpha},M^2_{\beta})$, where $M^2_{\alpha}$ is a sequence of $n$($=9$) complex values of $M^2$ appearing in the sum Eq.~(\ref{chi2}), can then be expressed in terms of the mentioned bin-covariance matrix $U$, Eq.~(\ref{covM}), in the following way:
\bes
\label{covMB}
\bea
(U_B)_{\alpha \beta} \equiv U_B(M^2_{\alpha},M^2_{\beta}) & = &
\langle \Delta {\rm Re} B(M^2_{\alpha}) \Delta {\rm Re} B(M^2_{\beta}) \rangle
\label{covMBa}
\\
& = & 
\sum_{j=1}^N \sum_{k=1}^N (\Delta \sigma)_j (\Delta \sigma)_k f(\sigma_j; M^2_{\alpha}) f(\sigma_k; M^2_{\beta}) U_{jk},
\label{covMBb}
\eea
\ees
where we used the notation
\be
\Delta {\rm Re} B(M^2) \equiv  {\rm Re} B(M^2) - \langle  {\rm Re} B(M^2) \rangle.
\label{DReB}
\ee
Here, $\langle {\rm Re} B(M_{\alpha}^2) \rangle$ is the central experimental value of (the real part of) the Borel sum rule at a scale $M^2=M^2_{\alpha}$, ($\alpha=1,\ldots,n$), i.e., it is the expression (\ref{ReBexp}) calculated with the central experimental spectral values $\langle \omega(\sigma_j) \rangle = \omega_{\rm exp}(\sigma_j)$. Square root of the diagonal elements of the above matrix
\be
\delta_B(M^2_{\alpha}) = \left(  U_B(M^2_{\alpha},M^2_{\alpha}) \right)^{1/2},
\label{dBM2}
\ee
which also appear in the sum $\chi^2$ of Eq.~(\ref{chi2}),
gives the experimental standard deviation for the Borel sum rule ${\rm Re} B(M^2)$ at a given scale $M^2_{\alpha}$.

Once we have the $n \times n$ covariance matrix $U_B$ ($n=9$ in our case) of the ${\rm Re} B_{\rm exp}(M^2_{\alpha})$ values, Eq.~(\ref{covMB}), and noting that the corresponding theoretical values ${\rm Re} B_{\rm th}(M^2_{\alpha})$ appearing in the sum $\chi^2$ of Eq.~(\ref{chi2}) depend on various parameters $p_j$ to be extracted [$p_1=a(\sm) =\alpha_s(\sm)/\pi$, $p_2=\langle O_4 \rangle$, etc.], we can proceed directly according to the formalism presented in Appendix of Ref.~\cite{Bo2011}, which gives the following result when the notations are adjusted to our considered case:
\bea
\langle \delta p_i \delta p_j \rangle &=& (A^{-1})_{i k} (A^{-1})_{j \ell} \frac{\partial {\rm Re} B_{\rm th}(M^2_{\alpha}; {\vec p})}{\partial p_k} \frac{\partial {\rm Re} B_{\rm th}(M^2_{\beta}; {\vec p})}{\partial p_{\ell}} \frac{U_B(M^2_{\alpha},M^2_{\beta})}{U_B(M^2_{\alpha},M^2_{\alpha}) U_B(M^2_{\beta},M^2_{\beta})},
\label{pipj} \eea
where ${\vec p} = (p_1, p_2,\ldots)$, and the $n \times n$ matrix $A$ is
\bea
A_{r s} & = & \frac{\partial {\rm Re} B_{\rm th}(M^2_{\gamma}; {\vec p})}{\partial p_r} \frac{\partial {\rm Re} B_{\rm th}(M^2_{\gamma}; {\vec p})}{\partial p_s}
\frac{1}{U_B(M^2_{\gamma},M^2_{\gamma})}.
\label{Am} \eea
The indices that appear repeatedly are summed over.

In the case of the V-channel, we have altogether 7 parameters to fit ($\alpha_s$, and six condensates $2 \langle O_{2 n} \rangle_{\rm V}$, $n=2,\ldots,7$). The evaluations described in this Appendix then involve inversion of a $7 \times 7$ matrix $A$, cf.~Eqs,~(\ref{pipj})-(\ref{Am}). This inversion turns out to be numerically unstable. For this reason, as an approximation we fixed the last parameter ($p_7= 2 \langle O_{14} \rangle_{\rm V}$), and reduced the problem to the inversion of a $6 \times 6$ matrix which turned out to be numerically stable. In this way, we obtained correlations $\langle \delta p_j \delta p_j \rangle^{1/2}$ (here no sum over $j$) which were close to those obtained when we, in addition, fixed even the penultimate parameter value $p_6 =  2 \langle O_{12} \rangle_{\rm V}$ (this is true at least for $i,j=1,2,3$).\footnote{E.g., in the FO case, the evaluation of $\delta a(\sigma_{\rm max})_{\rm exp} = \langle \delta p_1 \delta p_1 \rangle^{1/2}$ gave us $\pm 0.00171$ in the case of fixed $\langle O_{14} \rangle_{\rm V}$, and $\pm 0.00152$ in the case of both $\langle O_{14} \rangle_{\rm V}$ and $\langle O_{12} \rangle_{\rm V}$ fixed.} This indicates that fixing the last parameter ($p_7= 2 \langle O_{14} \rangle_{\rm V}$) gives a good approximation for the correlation values $\langle \delta p_i \delta p_j \rangle$, at least for $i,j=1,2,3,4$ [we recall that $p_1=\alpha_s(\sm)/\pi$], and thus for the experimental uncertainties $\delta p_j({\rm exp}) = \langle \delta p_j \delta p_j\rangle^{1/2}$. The experimental uncertainty of the last condensate $2 \langle O_{14} \rangle_{\rm V}$ was estimated by performing the fit via minimising the expression $\chi^2$ Eq.~(\ref{chi2}) with respect to the upper border of the experimental band\footnote{This means that in $\chi^2$ of Eq.~(\ref{chi2}) we replaced ${\rm Re}B(M^2_{\alpha}, \sm)$ by ${\rm Re}B(M^2_{\alpha}, \sm) + \delta_B(M^2_{\alpha})$.} and taking the resulting variation of the extracted value of $2 \langle O_{14} \rangle_{\rm V}$ as its experimental uncertainty: $10^3 \delta (2 \langle O_{14} \rangle_V) ({\rm exp}) \approx \pm 9.0 \ {\rm GeV}^{14}$, which is probably an overestimate.

 \end{document}